\documentclass[aps,pra,twocolumn,showpacs,superscriptaddress,groupedaddress,longbibliography]{revtex4-2}
\usepackage{latexsym,epsfig,amssymb,amsfonts,amsmath,graphicx,color,bm,times,amstext,epstopdf}
\usepackage[colorlinks]{hyperref}
\usepackage{mathdots}
\usepackage{yhmath}
\usepackage{cancel}
\usepackage{color}
\usepackage{siunitx}
\usepackage{array}
\usepackage{multirow}
\usepackage{amssymb}
\usepackage{textcomp,gensymb}
\usepackage{tabularx}
\usepackage{booktabs}
\usepackage{relsize}

\usepackage[misc,clock,geometry]{ifsym}

\usepackage{graphicx,color}          
\usepackage{epsfig}                  
\usepackage{wrapfig}
\usepackage{subfigure}
\usepackage{braket}

\usepackage[normalem]{ulem}

\begin{document}

\title{Quantum switch in the gravity of Earth}

\author{Nat\'alia S. M\'oller}\email{moller@fisica.ufmg.br}
\address{Departamento de F\'isica--ICEx, Universidade Federal de Minas Gerais, \\ CP702, 30161-970, Belo Horizonte, MG, Brazil}
\address{RCQI, Institute of Physics, Slovak Academy of Sciences, \\ Dúbravská Cesta 9, 84511 Bratislava, Slovakia}

\author{Bruna Sahdo}\email{brunasahdo@ufmg.br}
\address{Departamento de F\'isica--ICEx, Universidade Federal de Minas Gerais, \\ CP702, 30161-970, Belo Horizonte, MG, Brazil}

\author{Nelson Yokomizo}\email{yokomizo@fisica.ufmg.br}
\address{Departamento de F\'isica--ICEx, Universidade Federal de Minas Gerais, \\ CP702, 30161-970, Belo Horizonte, MG, Brazil}

\begin{abstract}
We introduce a protocol for a quantum switch in the gravitational field of a spherical mass and determine the time interval required for its realization in the gravity of Earth. One of the agents that perform operations with indefinite order is a quantum system in a path superposition state. Entanglement between its proper time and position is explored as a resource for the implementation of the quantum switch. The realization of the proposed protocol would probe the physical regime described by quantum mechanics on curved spacetimes, which has not yet been explored experimentally.\end{abstract}
	

\maketitle

\section{Introduction}

The quantum switch~\cite{Chiribella} is a task in which two noncommuting operations are realized on a target quantum system resulting in a quantum superposition of the orders they were applied. This task has already been implemented in optical tables~\cite{Procopio,Rubino,Goswami,Taddei} and provides a variety of advantages for quantum computation and communication~\cite{Araujo,Guerin,Wei,ChirDisc,Goswami2020,Guo2020}, quantum thermodynamics~\cite{Refrigerator_Vedral,RefrigExp} and quantum metrology~\cite{Zhao2020}. The idea of the quantum switch is formalized in the process matrix framework~\cite{ChirSupermap,Oreshkov}, a formulation of quantum mechanics in which the existence of a classical background of causal relations among events is not assumed~\cite{Hardy}. One can testify the indefinite order using a causal witness~\cite{Araujo2015,Rubino,Goswami} and via the Bell's theorem for temporal order~\cite{tbell,RubinoAgain}. The quantum switch is also employed in thought experiments in quantum gravity phenomenology~\cite{tbell}.

The quantum switch is achieved by means of a control quantum system C, whose state is entangled with the order of the operations (see review \cite{Goswami}). Suppose that $\ket{0}_c$ and $\ket{1}_c$ are orthogonal states of C, and $\mathcal{A},\mathcal{B}$ are noncommuting operations that can be applied by agents A and B to the target system. Let the composite system be such that, if C is in the state $\ket{0}_c$, the operation $\mathcal{A}$ is applied before the operation $\mathcal{B}$,
\begin{equation}
\ket{0}_c \ket{\psi} \mapsto \ket{0}_c \mathcal{B} \mathcal{A} \ket{\psi} \, ,
\end{equation}
and if C is in the state $\ket{1}_c$, the operations are applied in the opposite order
\begin{equation}
\ket{1}_c \ket{\psi} \mapsto \ket{1}_c \mathcal{A} \mathcal{B} \ket{\psi} \, .
\end{equation}
Preparing the control system in the state $(\ket{0} + \ket{1})/\sqrt{2}$, the target will evolve into a superposition of states obtained through the application of the operations $\mathcal{A}$ and $\mathcal{B}$ in switched orders,
\begin{equation}
\frac{\left( \ket{0}_c + \ket{1}_c \right)}{\sqrt{2}} \ket{\psi} \mapsto \frac{ \ket{0}_c \mathcal{B} \mathcal{A} \ket{\psi} + \ket{1}_c \mathcal{A} \mathcal{B} \ket{\psi} }{\sqrt{2}}\, .
\end{equation}
The order of the operations is then said to be indefinite.
	
Apart from applications in quantum computation and communication, processes with indefinite order appear in connection with foundational questions in quantum gravity. The hypothesis that the gravitational field can exist in a superposition of classical configurations has been explored as a possible path to quantum gravity phenomenology \cite{tbell,FORD1982238,Anastopoulos_2015,Bose, Marletto,Belechia_Wald_2018,Rovelli,Howl2020}. In particular, it was shown in~\cite{tbell} that, if the gravitational field of a mass in a superposition of distinct positions displays a corresponding superposition of classical configurations, then it can be used as the control of a quantum switch.  Observation of the quantum switch in this context would then testify that the gravitational field is in a superposition state. A quantum switch controlled by gravity is referred to as a gravitational quantum switch. The relation between the gravitational quantum switch and optical implementations in classical spacetime was discussed in ~\cite{Nikola}. A proposal for simulating the gravitational quantum switch using accelerated agents on Minkowski spacetime was described in~\cite{Rindler}. 

With the current technology, it is still a challenge to put a massive body in superposition for enough time to realize the gravitational quantum switch proposed in \cite{tbell}. In the present work, we introduce an alternative strategy for implementing the quantum switch in a gravitational system. It was argued in \cite{ZychRelQuantSup} that outcomes of a process where a localized system interacts with another system in a superposition of positions can be reproduced by a process where the first system is delocalized while the second is localized. Such a correspondence suggests that the gravitational quantum switch formulated in \cite{tbell} can be simulated in a new context where the mass is at a definite position, while the agents and target are delocalized. In this case, the quantum switch would be implemented by the dynamics of quantum systems on a definite curved spacetime, and could be explored as a tool for testing aspects of quantum mechanics on curved spacetimes, which is our main interest in this work.

We show that a protocol for a quantum switch analog to that of \cite{tbell} can indeed be formulated on the gravitational field of a central mass. Instead of attempting a direct translation of the protocol of \cite{tbell}, we present an alternative protocol that allows for a more efficient implementation. One of the agents is set in a superposition of positions on the curved classical spacetime describing the gravitational field of a central mass. We explore a setup in which the distances from the alternative positions of the agent to the central mass change over time. In contrast, static agents are considered in~\cite{tbell}. The proper times along the alternative paths are distinct, allowing the evolved state to display a superposition of proper times, which can be explored for the implementation of a quantum switch. With a careful choice of paths for the agent, the amount of time required to perform the quantum switch can be considerably decreased for a given experimental precision. Observation of such a quantum switch would confirm the production of a superposition of proper times by the distinct gravitational time dilations along the alternative paths of the agent, providing a test of quantum mechanics on curved spacetimes.

We first describe our general protocol and then compute its minimum duration on Earth's surface. Next we illustrate how the protocol could be implemented using few-level systems as agents. Then we discuss our results and their relation to previous protocols for the quantum switch.

\section{Quantum switch with entangled agents}

\begin{figure*}
\begin{center}
\hspace{-0.42cm}\includegraphics[scale=.37]{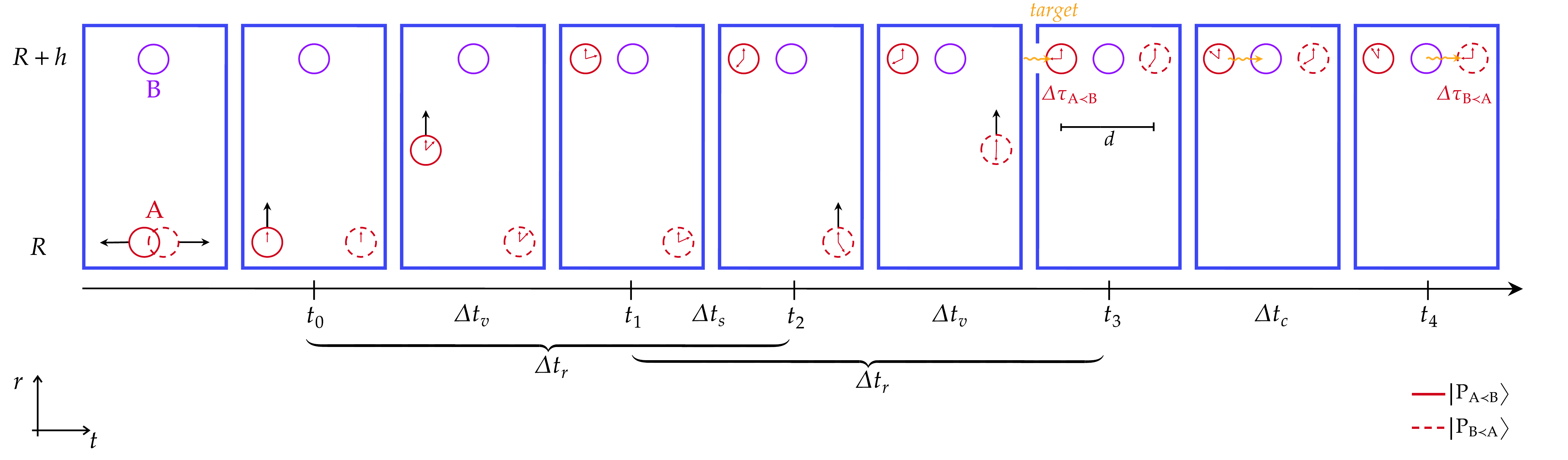}
\end{center}
\caption{Superposition of paths. The vertical axis represents the radius $r$ and the horizontal axis represents time. At $t=t_0$, the agent A is prepared in a superposition state $(|{\rm P}_{{\rm A}\prec {\rm B}}\rangle+|{\rm P}_{{\rm B}\prec {\rm A}}\rangle)/\sqrt{2}$. For the path $|{\rm P}_{{\rm A}\prec {\rm B}}\rangle$, A starts traveling up at $t_0$ to a height $h$ above the surface $r=R$. For the path $|{\rm P}_{{\rm B}\prec {\rm A}}\rangle$, it travels up in the same manner starting at $t_2$. The target system, traveling horizontally at height $h$, crosses $|{\rm P}_{{\rm A}\prec {\rm B}}\rangle$, meets agent B, and then crosses $|{\rm P}_{{\rm B}\prec {\rm A}}\rangle$.}
\label{LabsTraveling}
\end{figure*}

Consider a spherical body with mass $M$ and radius $R$. The gravitational field outside the body is described by the Schwarzschild metric,
\begin{equation}
ds^2 = -\left(1-\frac{R_S}{r}\right) c^2 dt^2 + \left(1-\frac{R_S}{r}\right)^{-1} dr^2 + r^2 d\Omega^2 \, ,
\label{eq:metric}
\end{equation}
where $d\Omega^2$ is the metric of the unit sphere and $R_S=2GM/c^2$ is the Schwarzschild radius. Our protocol for the quantum switch involves three quantum systems that can be manipulated in the vicinity of its surface $r=R$, which we call the agents A and B and the target system. By agents we mean systems that are able to interact with the target system, and thereby operate on its state. The three systems have nontrivial internal structures, which we will discuss in detail later. In this section, we first introduce the relevant features of quantum mechanics on curved spacetimes required for the description of the quantum agents in the Schwarzschild metric, and then present our proposed protocol for general noncommuting operations $\mathcal{A}$ and $\mathcal{B}$. A concrete implementation with a specific choice of operations is described in the next section.

\subsection{Quantum systems with internal degrees of freedom in a weak gravitational field}

The dynamics of quantum systems with internal degrees of freedom in a curved spacetime has been analyzed in several works in the regime of weak gravitational fields and motions that are slow in comparison with the speed of light (see \cite{Zych} and references therein). In particular, the case of a system with internal degrees of freedom in a weak gravitational field produced by a central mass is discussed in \cite{Zych2011,tbell}. The metric is then given by Eq.~\eqref{eq:metric}, with $R_S/r\ll1$, and is completely characterized by the gravitational potential
\begin{equation}
\Phi = - \frac{GM}{r} \, .
\label{eq:potential}
\end{equation}

Consider a system with internal degrees of freedom described in the absence of the gravitational potential by a Hilbert space $\mathcal{H}^{int}$. The same Hilbert space describes the internal degrees of system in the presence of the gravitational potential $\Phi$. The spatial configuration of the system is represented by a wavefunction $\psi(x) \in \mathcal{H}^{ext} \simeq L^2(\mathbb{R}^3;\mathbb{C})$. The full Hilbert space of the system is then $\mathcal{H} = \mathcal{H}^{int} \otimes \mathcal{H}^{ext}$. The Hamiltonian in the curved geometry can be written in terms of the Hamiltonian $H_{int}$ describing the system in the absence of a gravitational potential and the gravitational potential $\Phi$, as discussed in \cite{Zych,Zych2011,tbell}. It describes a modified Schr\"odinger equation that includes corrections due to the effect of gravitational time dilation in the evolution of the wavefunction and of the internal states.

The evolution of the internal degrees of freedom has a simple description when the wavefunction of the system is well localized at all times. Let  the support of $\psi(t,x)$ be restricted, for each $t$, to a finite region $V_{x(t)}$ around a point $x(t)$, within which the variation of the potential is negligible, $(\Phi(t,x')- \Phi(t,x))/ \Phi(x(t)) \ll 1$, $\forall x'\in V_{x(t)}$. In this case, the system has a well defined position at each time and a well defined proper time along its evolution, which for a slow motion is simply determined by
\begin{equation}
\frac{d \tau}{dt} = \sqrt{1 - \frac{R_S}{r}}  \, .
\end{equation}
We define a path $\rm{P}$ as a worldline $x(t)$ in spacetime. A path state $\ket{\rm{P}}$ is a wavefunction $\psi(t,x) \in \mathcal{H}^{ext}$ that is well localized at the event $x(t)$ of the path $\rm{P}$ for each instant of time $t$. The state at a given time $t$ is represented as $\ket{{\rm P};t}$. For localized states, the explicit form of the wavepacket is not relevant for the evolution of the internal state $\ket{\phi} \in \mathcal{H}^{int}$, which is determined by
\begin{equation}
i \frac{d}{d\tau_P} \ket{\phi}_{\rm P} = H_{int} \ket{\phi}_{\rm P} \, ,
\label{eq:internal-evolution-general}
\end{equation}
where $\tau_P$ is the proper time along the path $\rm{P}$, and $H_{int}$ is the Hamiltonian for the internal degrees of freedom in the absence of a gravitational potential. The state evolves with respect to the proper time as in the absence of the gravitational potential, but the proper time $\tau_P$ and the coordinate time $t$ are now distinct, being related by a factor describing the gravitational time dilation. Equivalently, the evolution can be written in terms of the coordinate time $t$ by
\begin{equation}
i \frac{d}{dt} \ket{\phi}_{\rm P} = H \ket{\phi}_{\rm P} \, ,
\end{equation}
with the Hamiltonian
\begin{equation}
H = \left. \frac{d\tau}{dt}\right|_{\rm P} H_{int} \simeq \left( 1+\frac{\Phi|_{\rm P}}{c^2} \right) H_{int} \, ,
\end{equation}
as presented in \cite{tbell}.

Consider a superposition of localized states of the form
\begin{equation}
\ket{\Psi} = \sum_i \ket{\rm{P}_i}\ket{\phi_i} \, .
\end{equation}
For brevity, we call such states path superposition states. For each path $\rm{P}_i$ in the superposition, the corresponding internal state $\ket{\phi_i} $ evolves according to Eq.~\eqref{eq:internal-evolution-general} with respect to the proper time along $\rm{P}_i$. Suppose that the initial state is separable, with $\ket{\phi_i}=\ket{\phi}$ at $t=t_0$. Time evolution will in general produce entanglement with respect to the bipartition  $\mathcal{H} = \mathcal{H}^{int} \otimes \mathcal{H}^{ext}$, as the internal state will evolve by distinct amounts of proper time along the distinct paths. In this sense, the proper times along the paths become entangled with the paths. On the other hand, if the proper times along the paths are the same at some instant of time $t$, then the time-evolved state will be separable again at such a time $t$, if the initial state is separable.

The dynamics of localized states in a curved spacetime provides a simple context for the analysis of gravitational effects in quantum systems. It is natural to expect the evolution of internal degrees of freedom to take place with respect to the proper time along the path followed by the localized state, as follows from the approach developed in \cite{Zych,Zych2011}. For path superposition states, the entanglement of the proper times with the spatial localization of the paths leads to new quantum effects that depend on the curvature of spacetime, as for instance a drop of visibility in quantum interferometric experiments \cite{Zych2011,Zych_2012} and a universal mechanism for decoherence in the position of composite particles \cite{Pikovski2015}.

In addition, models explored in recent works for the formulation of a phenomenology of the low-energy limit of quantum gravity \cite{Bose,Marletto,Rovelli,tbell} also rely on the validity of the quantum mechanics of nonrelativistic systems in curved spaces. In order to describe the evolution of a quantum system in a superposition of geometries, one must first be able to describe its evolution in each of the geometries in the superposition. In such works, it is assumed that an entanglement between proper times and paths takes place for each geometry. In the context of quantum gravity, the superposition of proper times in each geometry is combined with effects due to the superposition of geometries. The interferometric experiment proposed in \cite{Bose,Marletto} constitutes an example of a model involving superpositions of proper times in a superposition of geometries, as discussed in \cite{Rovelli}, as well as the gravitational quantum switch proposed in \cite{tbell}.

\subsection{Protocol for the quantum switch}
\label{sec:general-protocol}

Let us now describe our protocol for the quantum switch. The agents A and B and the target are quantum systems with internal degrees of freedom in the Schwarzschild metric \eqref{eq:metric}. We restrict to the weak-field regime $R_S/r \ll 1$. According to the discussion in the previous subsection, the Hilbert space of the agent A has the form $\mathcal{H}^{ext}_A \otimes \mathcal{H}^{int}_A$, where $\mathcal{H}^{int}_A $ describes its internal state and $\mathcal{H}^{ext}_A$ describes its position, and similarly for the agent B and target.

Our protocol involves a path superposition state for the agent A that includes two path states $|{\rm P}_{{\rm A}\prec {\rm B}}\rangle,  |{\rm P}_{{\rm B}\prec {\rm A}}\rangle \in \mathcal{H}^{ext}_A$, while B remains at a constant position at a height $h$ above the surface $r=R$ of the massive body that produces the gravitational field. The paths for A are represented in Fig.~\ref{LabsTraveling}. Both start from a common departure point at $r=R$ with the same angular position as that of agent B. Next, they separate horizontally in a symmetric manner up to a distance $d$. For the path ${\rm P}_{{\rm A}\prec {\rm B}}$, A starts traveling up at the instant $t_0$ until it reaches a point $X_{{\rm A}\prec {\rm B}}$ at $r=R+h$ at time $t_1$. Put $\Delta t_v=t_1-t_0$. A remains at this position afterwards. For ${\rm P}_{{\rm B}\prec {\rm A}}$, A also travels up to a point $X_{{\rm B}\prec {\rm A}}$ at $r=R+h$ in an interval $\Delta t_v$, but starting at a later time $t_2=t_1+\Delta t_s$. The target system, traveling horizontally, meets the point $X_{{\rm A}\prec {\rm B}}$ at $t_3$ and then travels towards $X_{{\rm B}\prec {\rm A}}$ in a time interval $\Delta t_c$. After that, the agent A is measured in a diagonal basis, as we will discuss in more detail later.

We consider a path superposition of the form
\begin{equation}
\ket{\Psi_{\rm A}} = \frac{1}{\sqrt{2}} \left(\ket{{\rm P}_{{\rm A}\prec {\rm B}}} \ket{\phi_{\rm A}}_{{\rm P}_{{\rm A}\prec {\rm B}}} + \ket{{\rm P}_{{\rm B}\prec {\rm A}}} \ket{\phi_{\rm A}}_{{\rm P}_{{\rm B}\prec {\rm A}}} \right) \, ,
\end{equation}
with
\begin{equation}
\ket{\phi_{\rm A};t_0}_{{\rm P}_{{\rm A}\prec {\rm B}}} = \ket{\phi_{\rm A};t_0}_{{\rm P}_{{\rm B}\prec {\rm A}}} \, ,
\end{equation}
i.e., we assume that the internal state of A is the same for both paths before they separate, in which case it remains the same while the paths remain at a common height, and that both paths are equally probable.

The agent A is configured to operate on the target at a specific instant $\tau^*$ in its proper time as indicated by an internal clock. This means that it is prepared in a state for which the probability of interacting with the target is considerable at $\tau^*$, but not at other times. The proper time of A must then be equal to $\tau^*$ when the target meets it for the interaction to take place. This must happen for both ${\rm P}_{{\rm A}\prec {\rm B}}$ and ${\rm P}_{{\rm B}\prec {\rm A}}$ for the interaction to occur regardless of the path taken. The agent B can interact with the target when their worldlines intersect. Under these conditions, the operations $\mathcal{A}$ and $\mathcal{B}$ are applied in distinct orders for each component $|{\rm P}_{{\rm A}\prec {\rm B}}\rangle$ or $|{\rm P}_{{\rm B}\prec {\rm A}}\rangle$ in the superposition of paths.

Let $\Delta\tau_{{\rm A}\prec {\rm B}}$ be the proper time along the path ${\rm P}_{{\rm A}\prec {\rm B}}$ from $t_0$ to the moment the target reaches it at $t_3$, and $\Delta\tau_{{\rm B}\prec {\rm A}}$ be the proper time along the path ${\rm P}_{{\rm B}\prec {\rm A}}$ from $t_0$ to the moment the target reaches it at $t_4$. Then the quantum switch will happen only if
\begin{equation}\label{1stCond}
\Delta\tau_{{\rm A}\prec {\rm B}}=\Delta\tau_{{\rm B}\prec {\rm A}} = \tau^* \, .
\end{equation}
Put $\Delta t_r\equiv\Delta t_v+\Delta t_s$. The interval $\Delta \tau_{{\rm A}\prec {\rm B}}$ has contributions from the time elapsed for A while it travels up and while it remains at radius $R + h$,
\begin{equation}\label{taua}
\Delta\tau_{{\rm A}\prec {\rm B}} 
=\Delta\tau_v +\sqrt{\left(1-\frac{R_S}{R+h}\right)} \ \Delta t_r \, ,
\end{equation}
where $\Delta \tau_v$ is the proper time elapsed for A while it travels up, corresponding to the coordinate time $\Delta t_v$. The interval $\Delta \tau_{{\rm B}\prec {\rm A}}$ includes contributions from the time elapsed for A while it stays at $r=R$, while it travels up, and while it waits the arrival of the target at $R+h$,
\begin{equation}\label{taub}
\Delta\tau_{{\rm B}\prec {\rm A}}
=\Delta\tau_v +\sqrt{\left(1-\frac{R_S}{R}\right)} \ \Delta t_r +\Delta\tau_{c} \, ,
\end{equation}
where $\Delta\tau_{c} = \sqrt{1-R_S/(R+h)} \ \Delta t_c$ and the proper time $\Delta \tau_v$ elapsed for the agent while it travels up is the same as in the other path. Substituting Eqs.~(\ref{taua}) and (\ref{taub}) into Eq.~(\ref{1stCond}), we find
\begin{equation} \label{main}
\left(
\sqrt{1-\frac{R_S}{R+h}}-\sqrt{1-\frac{R_S}{R}} \
\right)
\frac{\Delta t_r}{\Delta t_c} =\sqrt{1-\frac{R_S}{R+h}} \, .
\end{equation}
In the weak field regime, characterized by $R_S \ll R$, this equation reduces to
\begin{align}
\Delta t_r &= \frac{R}{R_S} \left( \frac{2R}{h} + 2  \right) \Delta t_c \nonumber \\
	& = \left( \frac{c^2}{g h} - \frac{c^2}{2} \frac{R_{0101}}{g^2} \right) \Delta t_c \, ,
	\label{eq:weak-field}
\end{align}
where $g=GM/R^2$ and $R_{0101}=-c^2 R_S/R^3$ is a component of the curvature tensor of the Schwarzschild metric (\ref{eq:metric}). The subleading term in the weak-field approximation is independent of $R_S/R$ and the following terms are proportional to higher powers of $R_S/R$. The first term in Eq.~(\ref{eq:weak-field}) depends on the acceleration of gravity at the radius where the experiment is performed. The second term describes the effect of curvature.

The parameter $\Delta t_r$ sets a time scale for the duration of the experiment. The total time of the protocol is $\Delta t_{\rm exp}\equiv t_4 - t_0$. For small $d$, such that $\Delta t_c \ll t_3-t_0$, this is well approximated by $t_3-t_0=\Delta t_r+\Delta t_v$. If the paths remain at distinct heights for a large amount of time, $\Delta t_v \ll \Delta t_s$, then we have $\Delta t_{\rm exp} \simeq \Delta t_r$. On the other hand, if $\Delta t_s= 0$, then $\Delta t_{\rm exp} \simeq 2 \Delta t_r$. In general, $\Delta t_{\rm exp} \sim \Delta t_r $.

Near the surface of the spherical mass, we can take $h\ll R$. The first term in Eq.~(\ref{eq:weak-field}) is then dominant. Considering the target to be a photon, we have $\Delta t_c\simeq d/c$. Under these approximations,
\begin{equation}\label{simple}
\Delta t_r \simeq \frac{cR^{2}d}{GMh} \, .
\end{equation}
We see that $\Delta t_r$ depends on two fundamental constants, $c$ and $G$; two properties of the massive body, $M$ and $R$; and two variables $d$ and $h$ that can be adjusted in the experiment. The duration of the experiment is minimized for the smallest possible distance $d$ between the paths and the largest possible height $h$. The distance $d$ in any implementation of the protocol will be limited by possible interactions between the agents and the precision of the clock. If the distance between the agents is so small that they can interact, their operations on the target will not be independent, as assumed. In addition, the clock must be sufficiently precise to resolve the time of flight $d/c$ of the photon between the paths of A. The height $h$ will be limited by the experimental capability of transporting A along its path-superposition state without decoherence.

Substituting the numerical values of $c,G$ and the radius $R_\odot$ and mass $M_\odot$ of the Earth in Eq.~(\ref{main}), we can estimate the duration of the experiment near the surface of Earth,
\begin{equation}\label{factorEarth}
\Delta t_{\rm exp} \sim 3 \times 10^7 \, \frac{d}{h} {\rm  s}\, .
\end{equation}
For an atomic clock with a precision of $10^{15} {\rm  Hz}$, for instance, the time of flight of the photon can be resolved for a distance of $0.3 \, \mu{\rm m}$. Setting $d = 0.3 \, \mu{\rm m}$ and $h=1 {\rm  m}$, we find $\Delta t_{\rm exp} \sim 9 {\rm  s}$.

We can also consider the case of a small mass. As in~\cite{tbell}, this example can be used to show that the effect does not require any physical quantity to be at the Planck scale to be observed. In this setting, it is natural to bring the departure point for the paths of A as close as possible to the mass, and we can take $h \gg R \gg R_S$. The second term in Eq.~(\ref{eq:weak-field}) is then dominant.  In this regime,
\begin{equation} \label{small-mass}
\Delta t_r \simeq \frac{cRd}{G M} \, .
\end{equation}

A well-known protocol for a gravitational quantum switch was previously formulated in the context of quantum gravity in \cite{tbell}. In that case, the agents and target move in a quantum state of the gravitational field produced by a mass in a superposition of positions.  In \cite{ZychRelQuantSup}, however, it is argued on general grounds that outcomes of a process in which a localized system interacts with a system in a superposition of positions can be reproduced by a process in which the first system is delocalized while the second system is localized. This suggests the possibility of simulating the gravitational quantum switch proposed in \cite{tbell} with delocalized quantum agents in the classical gravitational field of a central mass at a definite position. Our work was motivated by this correspondence, but it was not our purpose to exactly simulate the protocol of \cite{tbell}. Instead, we aimed to reproduce its relevant features using quantum agents in the Schwarzschild metric in order to obtain an efficient implementation of the quantum switch in this context, as a possible test of quantum mechanics on curved spacetimes. We can now compare the duration of our protocol with that expected for the protocol of \cite{tbell} under the correspondence proposed in \cite{ZychRelQuantSup}, as a means of testing its efficiency.

In the protocol of \cite{tbell}, for a small distance $d\ll R$ between the agents, the minimum proper time elapsed for the agents for the quantum switch to occur is $\tau^*= 2 r_b^2 c/GM$, where $r_b$ is the distance from the agents to the mass. Setting $r_b = R_\odot$, we find that $\tau^*$ is of the order of a year. This result can be compared with Eq.~\eqref{factorEarth}. The duration of the experiment is suppressed by a factor of $d/h$ using dynamical agents as described here.

For the small mass limit, a mass of $M = 0.1 \, \mu{\rm g}$ was considered in \cite{tbell}, with one agent at a distance of $1 {\rm  fm}$ and the other at a distance of $0.1\, \mu{\rm m}$ from the mass. The protocol for the Bell test using static agents explored in~\cite{tbell} would then take around $10 \, {\rm  h}$. Setting $R=1 \, {\rm  fm}$ in Eq.~\eqref{small-mass} and assuming $d=R$, we obtain $\Delta t_{\rm exp} \sim 5 \times 10^{-2} \, {\rm  s}$. In general, the duration of our protocol is of the order of one second if $Rd \sim 10^{-28} \, {\rm  m}^2$, and grows linearly with $Rd$.


\section{A model for the operations} 

We discussed the spacetime features required for the realization of the quantum switch in a Schwarzschild spacetime. We now explore possibilities for the operations performed by the agents. For concreteness, we consider a model involving a particular choice of quantum systems as agents and target.  The relevant features of the model are not restricted to this specific quantum system, however, which provides an illustration of a procedure that can be adapted to other systems of interest.

\subsection{Operations with indefinite order}

\begin{figure}[h]
	\begin{center}
		\includegraphics[scale=.23]{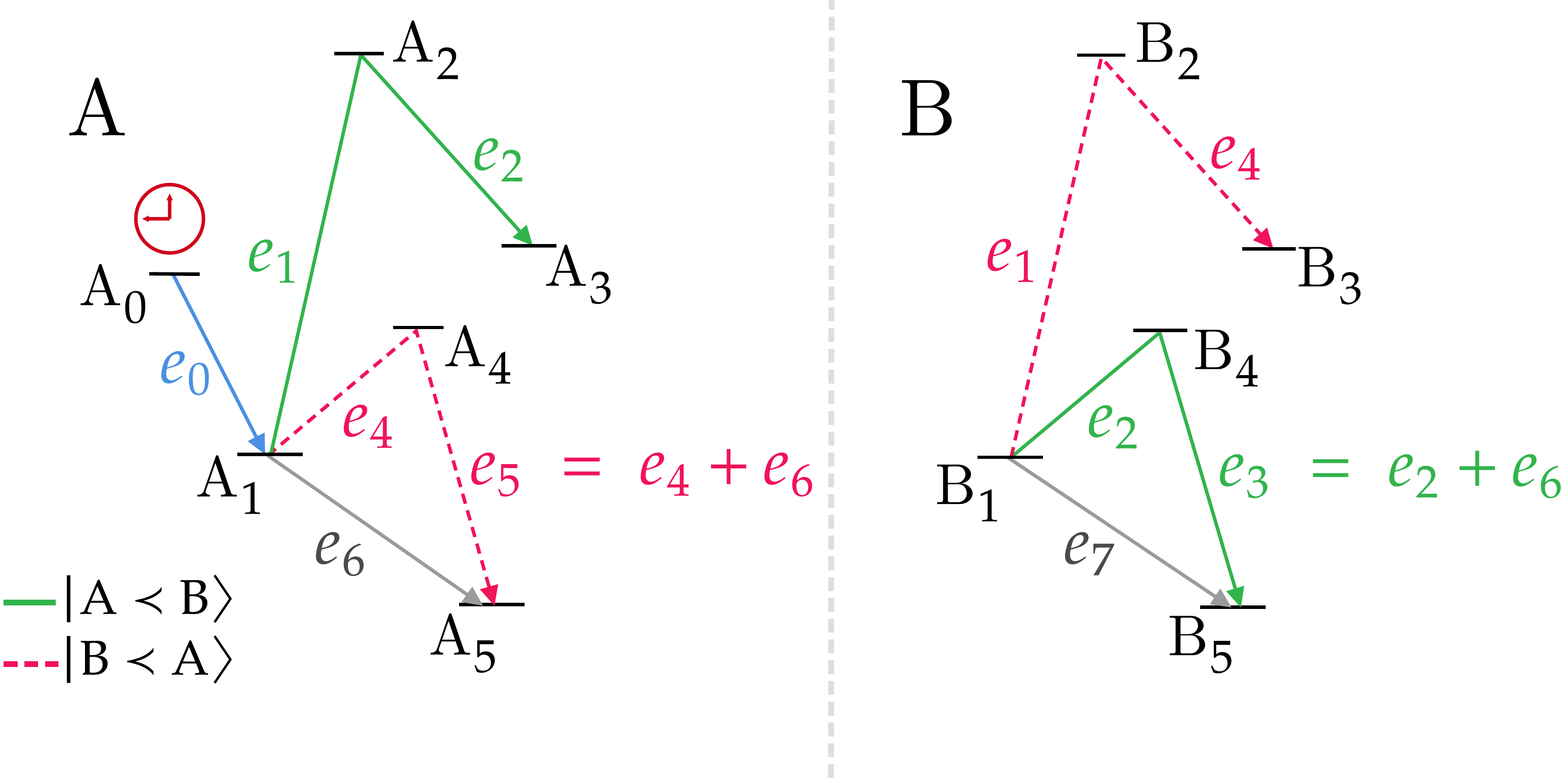}
		\caption{Energy levels of the agents A and B.}
        \label{model}
        \end{center}
\end{figure}

The internal Hilbert space of agent A has a subsystem $\mathcal{H}_\mathsmaller{\tiny \VarClock}$ which we call the trigger. It also includes a subsystem $\mathcal{H}_A$ with six energy levels $\ket{A_i}$, $i=0,\dots,5$. That is, $\mathcal{H}^{int}_A = \mathcal{H}_\mathsmaller{\tiny \VarClock} \otimes \mathcal{H}_A$. The trigger system will play the role of an internal clock for the agent A. The agent B is a system with internal degrees of freedom described by a Hilbert space $\mathcal{H}^{int}_B = \mathcal{H}_B$ with five energy levels $\ket{B_i}$, $i=1,\dots,5$. The energy level diagrams for $\mathcal{H}_A$ and $\mathcal{H}_B$ are represented in Fig.~\ref{model}. The labels $e_0$, $e_1$, etc, are energy differences between pairs of levels for the allowed transitions. We assume that the transitions are induced by the absorption and emission of photons.

The trigger is coupled to the six-level system $\mathcal{H}_A$. The time-evolution of internal degrees of freedom of a quantum system following a path $P$ in a curved spacetime is generated by an internal Hamiltonian $H_{int}$ evolving with respect to the proper time $\tau_P$ along $P$ \cite{tbell,Pikovski2015,Zych}, as described by Eq.~\eqref{eq:internal-evolution-general}. For the agent A, in particular,
\begin{equation}
i \frac{d}{d\tau_P} \ket{\phi_{\rm A}}_P = H_{int} \ket{\phi_{\rm A}}_P \, ,
\label{eq:internal-evolution}
\end{equation}
where $\ket{\phi_{\rm A}}_P \in \mathcal{H}_{\tiny \VarClock} \otimes \mathcal{H}_A$. For a path superposition state, the evolution of the internal state is described by Eq.~(\ref{eq:internal-evolution}) for each path in the superposition.  A complete description of the state $\ket{\Psi_{\rm A}}$ of A also includes its spatial location,
\begin{equation}
\ket{\Psi_{\rm A}} = \frac{1}{\sqrt{2}} \left(\ket{{\rm P}_{{\rm A}\prec {\rm B}}} \ket{\phi_{\rm A}}_{{\rm P}_{{\rm A}\prec {\rm B}}} + \ket{{\rm P}_{{\rm B}\prec {\rm A}}} \ket{\phi_{\rm A}}_{{\rm P}_{{\rm B}\prec {\rm A}}} \right) \, .
\label{eq:internal-evolution-superposition}
\end{equation}

At the beginning of the experiment, $\mathcal{H}_A$ is prepared in the state $\ket{A_0}$, which is stable in the absence of the trigger. The trigger is prepared in a state $\ket{\raisebox{-2pt}{\footnotesize \VarClock} \, ; \tau=0}$. We assume that a sharp transition from $\ket{A_0}$ to $\ket{A_1}$ is induced by the trigger after a proper time $\tau^*$ has elapsed for A since $t_0$, with $\tau^*$ given by Eq.~(\ref{1stCond}). Denoting the unitary evolution under $H_{int}$ by $U(\tau^*,0)$ and putting
\begin{equation}
\ket{\Psi_A;0} \equiv \ket{\raisebox{-2pt}{\footnotesize \VarClock}\, ;\tau=0} \ket{A_0} \, ,
\end{equation}
we require that
\begin{equation}
U(\tau,0) \ket{\Psi_{\rm A};0} \simeq \left\{ \begin{array}{ll}
	\ket{\raisebox{-2pt}{\footnotesize \VarClock}\, ;\tau} \ket{A_0}   \, , & {\rm for } \, \tau < \tau^* - \epsilon \, ,\\
	\ket{\raisebox{-2pt}{\footnotesize \VarClock}\, ;\tau^*} \ket{A_1}  \, , & {\rm for } \, \tau = \tau^* \, .
\end{array}
\right.
\label{eq:trigger-condition}
\end{equation}
We provide a concrete example of a unitary evolution satisfying the above properties in the Appendix, as an illustration of how such a trigger could be implemented. 

The trigger plays the role of a clock that at $\tau^*$ changes the state of A into a new state that can interact with the target. In other words, the instrument of A, represented by $\mathcal{H}_A$, is switched on by the trigger at $\tau^*$. It is in fact sufficient that the trigger generates a nonzero projection on $\ket{A_1}$. The agent B is prepared in the state $\ket{B_1}$ at the time $t_3+d/2c$. As B remains at a fixed position, an external clock can be used to prepare it in the required state at the scheduled time.

We require the levels $\ket{A_1}$ and $\ket{B_1}$ to have a small decay time satisfying $\Delta \tau_1 \ll d/c$ and $\epsilon \ll \Delta \tau_1$. A can then absorb a photon of energy $e_1$ or $e_4$ and get excited to the level $\ket{A_2}$ or $\ket{A_4}$ only if the photon arrives at $\tau^*$, within a time-window of approximately $\Delta \tau_1$. If no photon reaches the system at this time, it decays to the level $\ket{A_5}$ by emitting a photon of energy $e_6$, which testifies that A has not absorbed an incoming photon during the process. Similarly, the system $\mathcal{H}_B$ can be excited only at the coordinate time $t_3+d/2c$. If no photon reaches the system at this time, it decays to the level $\ket{A_5}$ by emitting a photon of energy $e_6$.

The experiment is designed so that, for $|{\rm P}_{{\rm A}\prec {\rm B}}\rangle$, a photon of energy $e_1$ meets A at $t_3$. If this photon is absorbed, A is excited to the level $\ket{A_2}$ and then rapidly decays to $\ket{A_3}$, emitting a photon of energy $e_2$. We represent such interaction as $\mathcal{A} \left(\ket{A_1} \ket{e_1} \right)=\ket{A_3} \ket{e_2}$. This step is understood as an operation $\mathcal{A}_\text{targ}$ on the photon state, i.e., $\mathcal{A}_\text{targ} |e_1\rangle=|e_2\rangle$. The emitted photon can be absorbed by B. When this does not happen, B decays to its ground state $\ket{B_5}$, emitting a photon of energy $e_7$ that testifies that the experiment was not completed. If the photon is absorbed, B is excited to the level $\ket{B_4}$ and quickly decays to $\ket{B_5}$ by emitting a photon of energy $e_3$. We represent such interaction as $\mathcal{B} \left(\ket{B_1} \ket{e_2} \right)=\ket{B_5} \ket{e_3}$. The operation $\mathcal{B}_\text{targ}$ on the photon state is $\mathcal{B}_\text{targ}|e_2\rangle=|e_3\rangle$. The operations are performed in the order $\mathcal{B} \mathcal{A}$. Defining
\begin{equation}
\ket{{\rm A}\prec{\rm B}} = \left|{\rm P}_{{\rm A}\prec{\rm B}}\right \rangle \ket{\raisebox{-2pt}{\footnotesize \VarClock}}_{{\rm A}\prec{\rm B}} \, ,
\end{equation}
where $\ket{\raisebox{-2pt}{\footnotesize \VarClock}}_{{\rm A}\prec{\rm B}}$ is the final state of the trigger, and introducing
\begin{equation}
\ket{\psi_1} = \ket{A_1} \ket{B_1} \ket{e_1} \, ,
\end{equation}
the final joint state of the agents and photon is 
\begin{equation}
\ket{{\rm A}\prec{\rm B}} \mathcal{B} \mathcal{A} \ket{\psi_1} = \ket{{\rm A}\prec{\rm B}} \ket{A_3}\ket{B_5} \ket{e_3} \, .
\end{equation}

For the path $|{\rm P}_{{\rm B}\prec {\rm A}}\rangle$, the sequence of events proceeds analogously, implementing operations $\mathcal{B} \left( \ket{B_1} \ket{e_1} \right)=\ket{B_3} \ket{e_4}$, with an action $\mathcal{B}_\text{targ} |e_1\rangle=|e_4\rangle$ on the target, and $\mathcal{A} \left( \ket{A_1} \ket{e_4} \right)=\ket{A_5} \ket{e_5}$, with an action $\mathcal{A}_\text{targ} |e_4\rangle=|e_5\rangle$ on the target, performed now in the switched order $\mathcal{A} \mathcal{B}$. In this case, the final joint state is
\begin{equation}
\ket{{\rm B}\prec{\rm A}} \mathcal{A} \mathcal{B} \ket{\psi_1} = \ket{{\rm B}\prec{\rm A}} \ket{A_5}\ket{B_3} \ket{e_5} \, ,
\end{equation}
where 
\begin{equation}
\ket{{\rm B}\prec{\rm A}} = \left|{\rm P}_{{\rm B}\prec{\rm A}}\right \rangle \ket{\raisebox{-2pt}{\footnotesize \VarClock}}_{{\rm B}\prec{\rm A}} \, .
\end{equation}

We assume the amplitude transitions for the processes in which the target interacts with both agents to be the same in the two paths. Then the final state of the system is
\begin{equation}
\frac{ \ket{{\rm A}\prec{\rm B}} \mathcal{B} \mathcal{A} \ket{\psi_1} + \ket{{\rm B}\prec{\rm A}} \mathcal{A} \mathcal{B} \ket{\psi_1} }{\sqrt{2}} \, .
\label{eq:switch-composite}
\end{equation}
A quantum switch is thus implemented, with the two alternative paths of the agent A playing the role of the control of the switch. As the agents are quantum systems, their interactions with the target are described by operators acting on $\mathcal{H}_A \otimes \mathcal{H}_B \otimes \mathcal{H}_\text{targ}$, where $\mathcal{H}_\text{targ}$ is the Hilbert space of the target, spanned by states $\ket{e_i}$. When the agents are classical, their operations are represented by operators on $\mathcal{H}_\text{targ}$.

Measuring the agents in a diagonal basis, the superposition of orders can be encoded in a superposition of target states. We define
\begin{align}
|{\rm F}_{{\rm A}\prec {\rm B}}\rangle &= \ket{{\rm A}\prec{\rm B}} \ket{A_3}\ket{B_5} \, , \\
|{\rm F}_{{\rm B}\prec {\rm A}}\rangle &= \ket{{\rm B}\prec{\rm A}} \ket{A_5}\ket{B_3}  \, .
\end{align}
Measuring the agents in the basis $|{\rm F}_{{\rm A}\prec {\rm B}}\rangle \pm|{\rm F}_{{\rm B}\prec {\rm A}}\rangle$ takes the photon to the state 
\begin{equation} \label{atomsfinal}
\frac{\mathcal{B}_\text{targ} \mathcal{A}_\text{targ} |e_1\rangle \pm \mathcal{A}_\text{targ} \mathcal{B}_\text{targ} |e_1\rangle}{\sqrt{2}}=\frac{|e_3\rangle \pm |e_5\rangle}{\sqrt{2}} \, ,
\end{equation}
and we obtain a superposition of the orders of the operations $\mathcal{A}_\text{targ}$ and $\mathcal{B}_\text{targ}$ on the target. The superposition of orders can then be verified by performing observations on the target system.

The measurement on the basis $|{\rm F}_{{\rm A}\prec {\rm B}}\rangle \pm|{\rm F}_{{\rm B}\prec {\rm A}}\rangle$ includes the measurement of the clock. This can be avoided by resynchronizing the clock states after the application of the operations, which would disentangle the clock from the rest of the system, allowing the measurement on the basis $|{\rm F}_{{\rm A}\prec {\rm B}}\rangle \pm|{\rm F}_{{\rm B}\prec {\rm A}}\rangle$ to be performed only on the path and few-level systems. This could be done by making A follow the paths of the protocol in a reversed way, similarly as done in~\cite{tbell}. Another possibility is to artificially synchronize the clock states by directly manipulating them, as done for instance in~\cite{Margalit}. 

We described the result of the operations of the agents A and B on an incoming photon of energy $e_1$, selecting runs of the experiment in which both agents absorbed some photon in the process. In this case, the operations $\mathcal{A} \mathcal{B}$ and $\mathcal{B} \mathcal{A}$ produce outgoing photon states of different energies, and the superposition of orders in the quantum switch leads to a superposition of final energies for the photon after a measurement in a diagonal basis, as described by Eq.~\eqref{atomsfinal}. As discussed in Sec.~\ref{sec:general-protocol}, this can happen only if there is a superposition of distinct proper times along the alternative paths of A, allowing the photon to cross the path ${\rm P}_{{\rm A}\prec {\rm B}}$ at $t_3$ and the path ${\rm P}_{{\rm B}\prec {\rm A}}$ at $t_4$ at the same proper time $\tau^*$ of A, as required for the operation $\mathcal{A}$ to be applied for both paths. Hence, the verification of the operation of the quantum switch for an incoming photon of energy $e_1$, as described by Eq.~\eqref{atomsfinal}, testifies to the superposition of proper times along the alternative paths.

The case of incoming photons of definite energies $e \neq e_1$ can be analyzed similarly. If the incoming photon has an energy $e \neq e_1$, then at most one agent can operate on it nontrivially. As a result, the final state of the system formed by the target and the few-level systems is the same for both paths after the application of the operations, making the switch of the order of operations trivial. For instance, an incoming photon with energy $e_4$ can interact nontrivially only with A, in which case the operation on the target is given by $\mathcal{A}_\text{targ} \ket{e_4}= \ket{e_5}$. For the interaction with B, we have $\mathcal{B}_\text{targ} \ket{e_4}=\ket{e_4}$ and $\mathcal{B}_\text{targ} \ket{e_5}=\ket{e_5}$. If A does not absorb the incoming photon, then it decays to its ground state, emitting a photon of energy $e_6$ that testifies that the interaction did not take place, and we can discard this run of the experiment.  The final state for the path ${\rm P}_{{\rm A}\prec {\rm B}}$ is
\begin{equation}
\ket{{\rm A}\prec{\rm B}}\ket{A_5}\ket{B_5} \ket{e_5}\, ,
\end{equation}
while the final state for ${\rm P}_{{\rm A}\prec {\rm B}}$ is obtained by replacing ${\rm A}\prec{\rm B}$ with ${\rm B}\prec{\rm A}$ in the expression above. The final state of the system is
\begin{equation}
\frac{\left( \ket{{\rm A}\prec{\rm B}} + \ket{{\rm B}\prec{\rm A}} \right)}{\sqrt{2}}\ket{A_5}\ket{B_5} \ket{e_5}
\end{equation}
The target is already disentangled from the rest of the system, and its final state is simply
\begin{equation}
\frac{\mathcal{B}_\text{targ} \mathcal{A}_\text{targ} |e_4\rangle + \mathcal{A}_\text{targ} \mathcal{B}_\text{targ} |e_4\rangle}{\sqrt{2}}= \ket{e_5} \, .
\label{eq:final-input-4}
\end{equation}
In this case, as for any state with definite energy $e \neq e_1$, it is not necessary to perform a measurement of the states of the agents before the measurement of the target, as the target disentangles from the rest of the system. The quantum switch is trivial, with both orders of operations producing the same result.

\subsection{Quantum switch for arbitrary input states}

In order to conceptually clarify the nature of the quantum switch implemented by the protocol, let us now discuss the case of a generic target state. As the quantum switch is nontrivial only for an incoming photon with initial energy $e_1$, our main purpose in analyzing the case of an arbitrary input target space is to discuss the general features of our proposal and clarify its relation to other implementations of the quantum switch.

In the examples previously considered, runs of the experiment were selected according to whether or not photons of energies $e_6$ or $e_7$ were produced. Accordingly, the formulation of the quantum switch for a generic input state involves a postselection of runs of the experiment, referring to the presence or not of the photons $e_6,e_7$. In addition, as the agents are quantum systems, each operation corresponds to an interaction between the target and the agent. In constrast, in implementations of the quantum switch with classical agents, the operations are represented directly on $\mathcal{H}_\text{targ}$.

Let the target system be described by the Hilbert space $\mathcal{H}_\text{targ}$ spanned by the states $\{\ket{e_i}; \, i=1,\dots,5\}$. This is sufficient for our purposes, since it includes all states involved in nontrivial interactions with the agents. The photons of energies $\ket{e_6}$ and $\ket{e_7}$ indicate whether or not each agent has absorbed some photon. Let $\mathcal{H}^A_d$ be the Hilbert space with basis $\{ \ket{0}_A,\ket{1}_A\}$, where the states $\ket{0}_A,\ket{1}_A$ describe configurations in which the photon $\ket{e_6}$ is absent or present, respectively. The Hilbert space $\mathcal{H}^B_d$ with basis $\{ \ket{0}_B,\ket{1}_B\}$ is defined analogously. Such photons are used to select runs of the experiment. There are four possible postselections, which we denote by $\zeta=0,1,2,3$ and which correspond to the situations where both photons, only $e_6$, only $e_7$, or none was emitted, respectively. We call the systems $\mathcal{H}^A_d$ and $\mathcal{H}^B_d$ the detectors of A and B.

Let us first consider the case of the path ${\rm P}_{{\rm A}\prec {\rm B}}$, for which the target first interacts with agent A and then with B. An incoming photon $\ket{e_i}$ can be absorbed or not by A. Let the amplitudes for these processes be
\begin{align}
& c_{iA} \, ,  & \text{if absorbed by A} \, , \nonumber \\
& d_{iA} = e^{i \delta_{iA}} \sqrt{|1-c_{iA}|^2} \, , & \text{if not absorbed by A} \, .
\end{align} 
The amplitudes $c_{iA}$ are nonzero only for $i=1,4$. The incoming photon may not be absorbed by A, but then be absorbed by B. The interaction of the incoming photon with B is described similarly in terms of amplitudes $c_{iB}$, which are nonzero for $i=1,2$. The photon can also interact nontrivially with both agents. This is possible only for an incoming photon of energy $e_1$. Let $f_{BA}$ be the amplitude for an incoming photon with energy $e_1$ that was scattered by A with energy $e_2$ to also be scattered by B, and the amplitude for the second scattering not to occur be $g_{BA}=e^{i \gamma_{BA}} \sqrt{1-|f_{BA}|^2}$.  The amplitude for double scattering is then $f_{BA} c_{1A}$. This exhausts all possible processes for ${\rm P}_{{\rm A}\prec {\rm B}}$.

By computing the  state that results from the interactions of A and B successively with a generic input state of the form 
\begin{equation}
\ket{\psi} = \ket{A_1} \ket{B_1} \sum_{i=1}^5 \alpha_i \ket{e_i} \, ,
\end{equation}
we find
\begin{multline}
U_B U_A \left( \ket{0}_A \ket{0}_B \ket{\psi} \right) \\
= \ket{0}_A \ket{0}_B \ket{\psi_{BA}} + \ket{0}_A \ket{1}_B\ket{\psi_{0A}} \\
+ \ket{1}_A \ket{0}_B \ket{\psi_{B0}} + \ket{1}_A \ket{1}_B \ket{\psi_{00}}  \, ,
\label{eq:BA-operations-1}
\end{multline}
where
\begin{align}
\ket{\psi_{BA}} &=  \alpha_1 c_{1A} f_{BA} \ket{A_3} \ket{B_5} \ket{e_3} \, ,\nonumber \\
\ket{\psi_{0A}} &=  \alpha_1 c_{1A} g_{BA} \ket{A_3} \ket{B_5} \ket{e_2} + \alpha_4 c_{4A} \ket{A_5} \ket{B_5} \ket{e_5} \, , \nonumber \\
\ket{\psi_{B0}} &= \alpha_1 d_{1A} c_{1B} \ket{A_5} \ket{B_3} \ket{e_4} + \alpha_2 c_{2B} \ket{A_5} \ket{B_5} \ket{e_3} \, , \nonumber \\
\ket{\psi_{00}} &= \sum_i \alpha_i d_{iA} d_{iB} \ket{A_5} \ket{B_5} \ket{e_i} \, .
\label{eq:BA-operations-2}
\end{align}

Let us now consider the path ${\rm P}_{{\rm B}\prec {\rm A}}$. Let $f_{AB}$ be the amplitude for an incoming photon with energy $e_1$ that was scattered by B with energy $e_4$ to be also scattered by A, and $g_{AB}= e^{i \gamma_{AB}} \sqrt{1-|f_{AB}|^2}$ be the amplitude for the second scattering not to occur. The amplitude for double scattering is then $f_{AB} c_{1B}$. After the application of both operations,
\begin{multline}
U_A U_B \left( \ket{0}_A \ket{0}_B \ket{\psi} \right) \\
= \ket{0}_A \ket{0}_B \ket{\psi_{AB}} + \ket{0}_A \ket{1}_B\ket{\psi_{A0}} \\
+ \ket{1}_A \ket{0}_B \ket{\psi_{0B}} + \ket{1}_A \ket{1}_B \ket{\psi_{00}}  \, ,
\label{eq:AB-operations-1}
\end{multline}
where
\begin{align}
\ket{\psi_{BA}} &=  \alpha_1 c_{1B} f_{AB} \ket{A_5} \ket{B_3} \ket{e_5} \, ,\nonumber \\
\ket{\psi_{0B}} &=  \alpha_1 c_{1B} g_{AB} \ket{A_5} \ket{B_3} \ket{e_4} + \alpha_2 c_{2B} \ket{A_5} \ket{B_5} \ket{e_3} \, , \nonumber \\
\ket{\psi_{A0}} &= \alpha_1 d_{1B} c_{1A} \ket{A_3} \ket{B_5} \ket{e_2} + \alpha_4 c_{4A} \ket{A_5} \ket{B_5} \ket{e_5} \, ,
\label{eq:AB-operations-2}
\end{align}
and $\ket{\psi_{00}}$ is given in Eq.~\eqref{eq:BA-operations-2}.

When the agent A is in the path superposition state \eqref{eq:internal-evolution-superposition}, the final state is given by
\begin{equation}
\frac{ \ket{{\rm A}\prec{\rm B}} U_B U_A  + \ket{{\rm B}\prec{\rm A}} U_A U_B }{\sqrt{2}} \left( \ket{0}_A \ket{0}_B \ket{\psi} \right) \, .
\end{equation}
The final state is a superposition of the states resulting from the interactions of agents A and B with the target in switched orders. This describes a quantum switch in an extended target space that includes the states of the detectors and of the few level systems. As the agents are quantum systems in our protocol, it is natural that their actions are described in a Hilbert space that includes the few-level systems. On the other hand, the detectors play a distinct role, allowing us to distinguish runs of the experiment where the target was scattered or not by the five- and six-level systems. We are interested in the case where the state of the detectors is measured after the interactions between the target and the agents. 

It turns out that for each possible outcome for the measurement of the detectors, the final state in $\mathcal{H}_A \otimes \mathcal{H}_B \otimes \mathcal{H}_\text{targ}$ is a superposition of states obtained by the application of the operations of A and B in switched orders, projected into the subspace associated with such an outcome. Concretely, let $\ket{\zeta}$ be the state of the detectors associated with the postselection $\zeta$ and $P^{(\zeta)}$ be the orthogonal projection on $\ket{\zeta}$. For instance, $\ket{\zeta=0} = \ket{1}_A \ket{1}_B$ and $P^{(0)} = \ket{1}_A \bra{1}_A \otimes \ket{1}_B \bra{1}_B$, and similarly for the other postselections. Let us introduce
\begin{align}
P^{(\zeta)} \left[ U_B U_A \left( \ket{0}_A \ket{0}_B \ket{\psi} \right) \right] &\equiv  \ket{\zeta} \mathcal{B}^{(\zeta)} \mathcal{A}^{(\zeta)} \ket{\psi} \, ,\nonumber \\
P^{(\zeta)} \left[ U_A U_B \left( \ket{0}_A \ket{0}_B \ket{\psi} \right) \right] &\equiv  \ket{\zeta} \mathcal{A}^{(\zeta)} \mathcal{B}^{(\zeta)} \ket{\psi} \, .
\label{eq:proj-zeta}
\end{align}
The explicit form of the states $\mathcal{B}^{(\zeta)} \mathcal{A}^{(\zeta)} \ket{\psi}$  and $\mathcal{A}^{(\zeta)} \mathcal{B}^{(\zeta)} \ket{\psi}$ can be directly extracted from Eqs.~\eqref{eq:BA-operations-1} and \eqref{eq:AB-operations-1} using the definitions \eqref{eq:proj-zeta}. We find that, for each postselection, the final state of the system formed by the few-level systems and target assumes the form:
\begin{equation}
\ket{{\rm A}\prec{\rm B}} \mathcal{B}^{(\zeta)} \mathcal{A}^{(\zeta)}  \ket{\psi} + \ket{{\rm B}\prec{\rm A}} \mathcal{A}^{(\zeta)} \mathcal{B}^{(\zeta)}  \ket{\psi} \, .
\label{eq:before-diagonal}
\end{equation}
The result is a quantum switch in $\mathcal{H}_A \otimes \mathcal{H}_B \otimes \mathcal{H}_\text{targ}$ controlled by the path of A. If the input target state has a vanishing projection in the state $\ket{e_1}$, i.e. $\alpha_1=0$, both orders of operations produce the same result, and the switch is trivial. The postselection $\zeta=3$, for which no detector clicked, selects states with $\alpha_1 \neq 0$, and the final state is independent of the other components $\alpha_i$, $i=2,\dots,5$, of the input state. After normalization, it is then always given by Eq.~\eqref{eq:switch-composite}. This postselection thus allows us to restrict to the nontrivial part of the quantum switch.

For any input target state, a measurement in the diagonal basis $\ket{{\rm A}\prec{\rm B}}\pm \ket{{\rm B}\prec{\rm A}}$ can be performed on the final state \eqref{eq:before-diagonal} in order to transfer the superposition of orders into a superposition of states in $\mathcal{H}_A \otimes \mathcal{H}_B \otimes \mathcal{H}_\text{targ}$, resulting in a state of the form
\begin{equation}
\mathcal{B}^{(\zeta)} \mathcal{A}^{(\zeta)}  \ket{\psi} \pm  \mathcal{A}^{(\zeta)} \mathcal{B}^{(\zeta)}  \ket{\psi} \, .
\end{equation}
In special cases, one can alternatively perform a measurement in a diagonal basis that includes the states of the few-level systems in order to transfer the superposition of orders into a superposition of target states, as in the cases previously discussed where the input state is a basis vector $\ket{e_i}$. This is convenient, in particular, for the most relevant case, to our purposes, of an input state $\ket{e_1}$.


\section{Discussion}

We have introduced a protocol for the implementation of a quantum switch in a gravitational system. Instead of considering classical agents A and B operating on a target moving on a superposition state of the gravitational field, we allowed the agents to be quantum systems, with A in a path superposition state, on a fixed curved background geometry produced by a central mass. Proper times along distinct paths are then entangled with the paths. With a careful choice of paths, we constructed a protocol that mirrors the relevant features of the protocol for a gravitational quantum switch proposed in \cite{tbell}. A test of Bell's inequality for temporal order can be implemented with two entangled copies of the agents and target.

In our protocol, the order of the operations is not entangled with the spacetime metric, which is classical, but with paths of a quantum system in this fixed curved background. Its realization would then consist of a test of quantum mechanics on curved spacetimes \cite{Pikovski2015,Lammerzahl:1995zz,Kiefer1991}, the limit of quantum field theory on curved spacetimes with negligible particle creation or annihilation and nonrelativistic speeds. This physical regime has not yet been probed experimentally, and our results provide a tool for testing the frequently adopted formulation of time-evolution on a curved spacetime leading to Eq.~(\ref{eq:internal-evolution-superposition}).

The quantum switch has been realized experimentally in non-gravitational systems~\cite{ReviewExp}. In such experiments, one does not keep track of the proper times at which agents perform their operations. If $\mathcal{A}$ is applied at distinct proper times of A for the orders $\mathcal{A} \mathcal{B}$ or $\mathcal{B} \mathcal{A}$, then measuring the time of the operation would in fact destroy the superposition of the order of operations. In our case, one agent is assumed to be equipped with an internal clock and apply its operation only at a prescribed time. This ensures that the influence of gravity on proper times along the distinct paths is the underlying effect allowing for the superposition of orders to occur.

Experiments that attest quantum phenomena due to the gravity of Earth in the Newtonian regime have already been made~\cite{Collela,Strelkov}. Time dilation is a dominant general relativistic correction to Newtonian gravity, and can be observed even for a height difference of $1 \, {\rm m}$~\cite{Chou2010}. A natural next step would be the exploration of superposition and entanglement of quantum clocks taking time dilation into account, an issue that has been theoretically explored \cite{Zych2011,Zych_2012,Terno2015,Rivera_Tapia_2020,RelHOM,Roura} and simulated with magnetic fields~\cite{Margalit}, but for which an experimental test with the gravitational field is still missing. With the progress on techniques for manipulating path superposition states at macroscopic scales~\cite{Dickerson2013,Kovachy2015,Hannover}, such tests might provide a path for the observation of quantum effects in gravitational systems, and our results include the quantum switch in a list of possible experiments aimed in this direction.

\begin{acknowledgments} 

The authors thank Raphael Drumond, Artur Matoso, Daniele Telles, Davi Barros, Gilberto Borges, Mariana Barros, Mario Mazzoni, Leonardo Neves, Pablo Saldanha, Sheilla Oliveira, {\v C}aslav Brukner, Aleksandra Dimi\'c, Marko Milivojevi\'c, and Dragoljub Go{\v c}anin for useful discussions. We also thank Marko Vojinovi\'c and Nikola Paunkovi\'c for useful comments on the manuscript.

B.S. acknowledges financial support from Coordena\c{c}\~ao de Aperfei\c{c}oamento de Pessoal de N\'{i}vel Superior (CAPES), Programa de Excelência Acadêmica (PROEX), under the Process No.~88887.495426/2020-00.
N.Y. acknowledges financial support from the Conselho Nacional de Desenvolvimento Cient\'{i}fico e Tecnol\'{o}gico (CNPq) under Grant No.~306744/2018-0. 
N.S.M. thanks VI Quantum Information School -- Paraty 2017 for the introduction on the topic of indefinite causal order. N.S.M. acknowledges support from the Coordena\c{c}\~ao de Aperfei\c{c}oamento de Pessoal de N\'{i}vel Superior (CAPES), Programa Institucional de Internacionaliza\c{c}\~ao CAPES-PrInt, under the Process No. 88887.474432/2020-00, and from Grant No. 61466 from the John Templeton Foundation, as part of the “The Quantum Information Structure of Spacetime (QISS)” Project (qiss.fr) and through Project No. VEGA 2/0161/19 (HOQIP). The opinions expressed in this publication are those of the author(s) and do not necessarily reflect the views of the John Templeton Foundation.

\end{acknowledgments}

\

N.S.M. and B.S. contributed equally to this work.


\appendix*

\section{A model for the trigger}

Let us describe a concrete implementation of a trigger satisfying the condition \eqref{eq:trigger-condition} discussed in the main text. We model the trigger as a harmonic oscillator $\mathcal{H}_\mathsmaller{\tiny \VarClock}=L^2(\mathbb{R})$ with free Hamiltonian
\begin{equation}
H_0 = - \frac{\hbar^2}{2m} p^2 + \frac{m \omega^2}{2} q^2
\label{eq:free-hamiltonian-trigger}
\end{equation}
and period
\begin{equation}
T = \frac{2 \pi}{\omega} = 4 \tau^* \, .
\label{eq:alarm-time}
\end{equation}
In the proposed protocol, the trigger plays the role of a clock that is programmed to change the state of the six-level system $\mathcal{H}_A$ of the agent A from $\ket{A_0}$ to $\ket{A_1}$ at a time $\tau^*$. We store the information about the predetermined time $\tau^*$ in the period $T$ of the oscillator through Eq.~\eqref{eq:alarm-time}.

The interaction of the oscillator with the system $\mathcal{H}_A$ is described by a Hamiltonian $H_{int}$ that is nonzero only on the subspace generated by the relevant states $\ket{A_0},\ket{A_1} \in \mathcal{H}_A$. Let $\sigma_x$ be the first Pauli matrix on this subspace,
\begin{equation}
\sigma_x = \begin{pmatrix}
0 & 1 \\ 1 & 0 
\end{pmatrix} \, .
\end{equation}
The interaction Hamiltonian is defined as
\begin{equation}
H_{int} = P_\Delta \otimes V_0 \, \sigma_x   \, , \qquad V_0 >0 \, ,
\end{equation}
where $P_\Delta$ is the orthogonal projection onto the region $x \in [0,\Delta]$, which acts on the wavefunction $\phi(x)$ of the oscillator according to
\begin{equation}
P_\Delta \phi(x) = \begin{cases}
\phi(x)  \, , & \text{if } x \in [0,\Delta] \\
0 \, , & \text{else} \, .
\end{cases}
\end{equation}
The interaction is nontrivial only when the oscillator is in the region $[0,\Delta]$, which we call the interaction zone. The full Hamiltonian of the system is
\begin{equation}
H = H_0 \otimes \bm{1}+ H_{int} \, .
\end{equation}

Let $\ket{\pm}$ be the eigenvectors of $\sigma_x$ with eigenvalues $\pm 1$,
\begin{equation}
\sigma_x \ket{\pm} = \pm \ket{\pm} \, .
\end{equation}
Then,
\begin{align*}
( H_0 \otimes \bm{1} + H_{int}) (\ket{\phi} \ket{+}) = \left[ (H_0 + V_0 P_\Delta) \ket{\phi}\right] \ket{+} \, , \\ 
( H_0 \otimes \bm{1} + H_{int}) (\ket{\phi} \ket{-}) = \left[ (H_0 - V_0 P_\Delta) \ket{\phi}\right] \ket{-} \, .
\end{align*}
Hence, for $\ket{\pm}$, the wavefunction evolves under a Hamiltonian
\[
H_0 + V_\Delta^\pm \, , \quad V_\Delta^\pm = \begin{cases}
\pm V_0 \, , & \text{if } x \in [0,\Delta] \, , \\
0 \, , & \text{else} \, .
\end{cases}
\]
For $\ket{+}$, the oscillator encounters a potential barrier; for $\ket{-}$, it encounters a potential well. The general case is a superposition of these situations.

Let $\ket{\alpha}$ be a coherent state of the trigger, $a \ket{\alpha}=\alpha \ket{\alpha}$, where $a$ is the annihilation operator of the harmonic oscillator. The trigger is prepared in a coherent state $\ket{\raisebox{-2pt}{\footnotesize \VarClock}\, ;\tau=0}=\ket{\alpha_0}$, where
\begin{equation}
\alpha_0 = \frac{A}{\sqrt{2} \sigma} \, , \qquad \sigma = \sqrt{\frac{\hbar}{m\omega}} \, , \qquad A = \frac{2 \Delta V_0}{\pi \hbar \omega} \, .
\label{eq:parameters}
\end{equation}
The parameter $\sigma$ is the width of the wavepacket. The coherent state describes a configuration of maximum positive displacement for an oscillation of amplitude $A$. The system $\mathcal{H}_A$ is prepared in the state $\ket{A_0}$ at $\tau=0$. The state of the composite system will be represented by $\ket{\Phi(\tau)}$.

We assume that $A \gg \Delta \gg \sigma$. The inequality $\Delta \gg \sigma$ means that the width of the wavepacket is much smaller than the width of the interaction region, i.e., that the state of the oscillator is well localized with respect to the potential step. The condition $A \gg \Delta$ means that the oscillator is initially far away from the interaction zone. Its evolution is thus initially determined by the free Hamiltonian $H_0$. As a result, it remains a coherent state $\ket{\raisebox{-2pt}{\footnotesize \VarClock}\, ;\tau}=\ket{\alpha(\tau)}$, where $\alpha(\tau)=\alpha_0 e^{-i\omega \tau}$, until it reaches the interaction zone. As the average position of such a coherent state is simply
\begin{equation}
\langle x \rangle = A \cos \omega t \, ,
\end{equation}
the wavepacket reaches the boundary of the interaction zone at $x=\Delta$ with a speed $v \sim \omega A$ after a time $\Delta \tau \simeq \tau^* - \epsilon$, where $\epsilon \sim \Delta/v$.  From $A, \Delta \gg \sigma$ and Eq.~\eqref{eq:parameters}, we also find that the energy of the wavepacket satisfies
\begin{equation}
\frac{m\omega^2 A^2}{2} \gg V_0 \, ,
\end{equation}
i.e., the energy of the wavepacket is much larger than the potential step $V_0$. We can then neglect the reflection of the wavepacket by the potential step and adopt the approximation of perfect transmission.

For $\tau < \tau^* - \epsilon$, the interaction Hamiltonian is negligible, since the wavepacket is outside the interaction zone:
\[
H_{int} \left( \ket{\alpha(t)} \ket{\chi} \right) = 0 \, , \quad \text{for } \, \tau<\tau^* -\epsilon \, ,
\]
for any $\ket{\chi} \in \mathcal{H}_A$. Therefore, the six-level system remains at the initial state $\ket{A_0}$ for $\tau<\tau^* -\epsilon$, and the first condition in Eq.~\eqref{eq:trigger-condition} is satisfied. The wavepacket then enters the interaction zone and crosses it in a time interval $\Delta \tau \simeq \Delta/v=\epsilon$. During this time, we have $P_\Delta \ket{\raisebox{-2pt}{\footnotesize \VarClock}\, ;\tau}\simeq \ket{\raisebox{-2pt}{\footnotesize \VarClock}\, ;\tau}$, so that 
\begin{align*}
H \ket{\Phi} &=  (\bm{1} \otimes H_0 + H_{int}) \ket{\Phi} \nonumber \\
	&\simeq (\bm{1} \otimes H_0 + V_0 \, \sigma_x \otimes \bm{1}) \ket{\Phi} \, ,\quad \text{for } \, \tau \in [\tau^*-\epsilon, \tau^*] \, .
\end{align*}
The time evolution generated by this Hamiltonian can be integrated exactly. We find that, for $\tau \in [\tau^*-\epsilon, \tau^*]$, the system evolves according to
\begin{equation}
\ket{\Phi(\tau)} = \left\{ e^{-iV_0 \sigma_x [\tau-(\tau^*-\epsilon)]/ \hbar} \ket{A_0} \right\} \ket{\alpha(\tau)} \, .
\end{equation}
At $\tau=\tau^*$, when the wavepacket reaches the opposite boundary of the interaction zone at $x=0$, we have
\begin{equation}
\ket{\Phi(\tau^*)} = \ket{A_1} \ket{\alpha(\tau^*)} \, ,
\end{equation}
and the second condition in Eq.~\eqref{eq:trigger-condition} is also satisfied, showing that the model described in this Appendix satisfies the properties required of the trigger.

Let us note that the scattering of a Gaussian wavepacket by a potential step is studied in \cite{Bernardini,Norsen} in the regime where the wavepacket has a small width in comparison with the step. The evolution of the wavepacket in this regime, which is usually not discussed in basic textbooks, can be described as a process involving multiple instantaneous scatterings with the boundaries of the potential step. The incident wavepacket branches into two wavepackets when it meets the potential step, with one branch corresponding to the transmitted wave and the other describing the reflected component. The transmitted wavepacket branches again into two new wavepackets when it meets the opposite boundary of the potential step, and so on. The initial wavepacket evolves in this manner into an infinite train of successive wavepackets, both reflected and transmitted, of progressively smaller amplitudes, as described in \cite{Bernardini}. We adopted the approximation of perfect transmission, valid for large energies, for the calculations above, so that the wavepacket crosses the interaction zone without branching into a superpositions of localized states. While it crosses the potential $V_0 \sigma_x$ in the interaction zone $[0,\Delta]$, it induces a rotation of the state of $\mathcal{H}_A$, and the parameters of the model can be adjusted so that the initial state $\ket{A_0}$ evolves into the state $\ket{A_1}$, as required.

\bibliography{ref}

\begin{thebibliography}{50}%
\makeatletter
\providecommand \@ifxundefined [1]{%
 \@ifx{#1\undefined}
}%
\providecommand \@ifnum [1]{%
 \ifnum #1\expandafter \@firstoftwo
 \else \expandafter \@secondoftwo
 \fi
}%
\providecommand \@ifx [1]{%
 \ifx #1\expandafter \@firstoftwo
 \else \expandafter \@secondoftwo
 \fi
}%
\providecommand \natexlab [1]{#1}%
\providecommand \enquote  [1]{``#1''}%
\providecommand \bibnamefont  [1]{#1}%
\providecommand \bibfnamefont [1]{#1}%
\providecommand \citenamefont [1]{#1}%
\providecommand \href@noop [0]{\@secondoftwo}%
\providecommand \href [0]{\begingroup \@sanitize@url \@href}%
\providecommand \@href[1]{\@@startlink{#1}\@@href}%
\providecommand \@@href[1]{\endgroup#1\@@endlink}%
\providecommand \@sanitize@url [0]{\catcode `\\12\catcode `\$12\catcode
  `\&12\catcode `\#12\catcode `\^12\catcode `\_12\catcode `\%12\relax}%
\providecommand \@@startlink[1]{}%
\providecommand \@@endlink[0]{}%
\providecommand \url  [0]{\begingroup\@sanitize@url \@url }%
\providecommand \@url [1]{\endgroup\@href {#1}{\urlprefix }}%
\providecommand \urlprefix  [0]{URL }%
\providecommand \Eprint [0]{\href }%
\providecommand \doibase [0]{https://doi.org/}%
\providecommand \selectlanguage [0]{\@gobble}%
\providecommand \bibinfo  [0]{\@secondoftwo}%
\providecommand \bibfield  [0]{\@secondoftwo}%
\providecommand \translation [1]{[#1]}%
\providecommand \BibitemOpen [0]{}%
\providecommand \bibitemStop [0]{}%
\providecommand \bibitemNoStop [0]{.\EOS\space}%
\providecommand \EOS [0]{\spacefactor3000\relax}%
\providecommand \BibitemShut  [1]{\csname bibitem#1\endcsname}%
\let\auto@bib@innerbib\@empty
\bibitem [{\citenamefont {Chiribella}\ \emph {et~al.}(2013)\citenamefont
  {Chiribella}, \citenamefont {D'Ariano}, \citenamefont {Perinotti},\ and\
  \citenamefont {Valiron}}]{Chiribella}%
  \BibitemOpen
  \bibfield  {author} {\bibinfo {author} {\bibfnamefont {G.}~\bibnamefont
  {Chiribella}}, \bibinfo {author} {\bibfnamefont {G.~M.}\ \bibnamefont
  {D'Ariano}}, \bibinfo {author} {\bibfnamefont {P.}~\bibnamefont
  {Perinotti}},\ and\ \bibinfo {author} {\bibfnamefont {B.}~\bibnamefont
  {Valiron}},\ }\bibfield  {title} {\bibinfo {title} {Quantum computations
  without definite causal structure},\ }\href
  {https://doi.org/10.1103/PhysRevA.88.022318} {\bibfield  {journal} {\bibinfo
  {journal} {Phys. Rev. A}\ }\textbf {\bibinfo {volume} {88}},\ \bibinfo
  {pages} {022318} (\bibinfo {year} {2013})}\BibitemShut {NoStop}%
\bibitem [{\citenamefont {Procopio}\ \emph {et~al.}(2015)\citenamefont
  {Procopio}, \citenamefont {Moqanaki}, \citenamefont {Araújo}, \citenamefont
  {Costa}, \citenamefont {Alonso~Calafell}, \citenamefont {Dowd}, \citenamefont
  {Hamel}, \citenamefont {Rozema}, \citenamefont {Brukner},\ and\ \citenamefont
  {Walther}}]{Procopio}%
  \BibitemOpen
  \bibfield  {author} {\bibinfo {author} {\bibfnamefont {L.~M.}\ \bibnamefont
  {Procopio}}, \bibinfo {author} {\bibfnamefont {A.}~\bibnamefont {Moqanaki}},
  \bibinfo {author} {\bibfnamefont {M.}~\bibnamefont {Araújo}}, \bibinfo
  {author} {\bibfnamefont {F.}~\bibnamefont {Costa}}, \bibinfo {author}
  {\bibfnamefont {I.}~\bibnamefont {Alonso~Calafell}}, \bibinfo {author}
  {\bibfnamefont {E.~G.}\ \bibnamefont {Dowd}}, \bibinfo {author}
  {\bibfnamefont {D.~R.}\ \bibnamefont {Hamel}}, \bibinfo {author}
  {\bibfnamefont {L.~A.}\ \bibnamefont {Rozema}}, \bibinfo {author}
  {\bibfnamefont {{\v C}.}~\bibnamefont {Brukner}},\ and\ \bibinfo {author}
  {\bibfnamefont {P.}~\bibnamefont {Walther}},\ }\bibfield  {title} {\bibinfo
  {title} {Experimental superposition of orders of quantum gates},\ }\href
  {https://doi.org/10.1038/ncomms8913} {\bibfield  {journal} {\bibinfo
  {journal} {Nat. Commun.}\ }\textbf {\bibinfo {volume} {6}},\ \bibinfo {pages}
  {7913} (\bibinfo {year} {2015})}\BibitemShut {NoStop}%
\bibitem [{\citenamefont {Rubino}\ \emph {et~al.}(2017)\citenamefont {Rubino},
  \citenamefont {Rozema}, \citenamefont {Feix}, \citenamefont {Araújo},
  \citenamefont {Zeuner}, \citenamefont {Procopio}, \citenamefont {Brukner},\
  and\ \citenamefont {Walther}}]{Rubino}%
  \BibitemOpen
  \bibfield  {author} {\bibinfo {author} {\bibfnamefont {G.}~\bibnamefont
  {Rubino}}, \bibinfo {author} {\bibfnamefont {L.~A.}\ \bibnamefont {Rozema}},
  \bibinfo {author} {\bibfnamefont {A.}~\bibnamefont {Feix}}, \bibinfo {author}
  {\bibfnamefont {M.}~\bibnamefont {Araújo}}, \bibinfo {author} {\bibfnamefont
  {J.~M.}\ \bibnamefont {Zeuner}}, \bibinfo {author} {\bibfnamefont {L.~M.}\
  \bibnamefont {Procopio}}, \bibinfo {author} {\bibfnamefont {{\v
  C}.}~\bibnamefont {Brukner}},\ and\ \bibinfo {author} {\bibfnamefont
  {P.}~\bibnamefont {Walther}},\ }\bibfield  {title} {\bibinfo {title}
  {Experimental verification of an indefinite causal order},\ }\href
  {https://doi.org/10.1126/sciadv.1602589} {\bibfield  {journal} {\bibinfo
  {journal} {Science Advances}\ }\textbf {\bibinfo {volume} {3}},\ \bibinfo
  {pages} {e1602589} (\bibinfo {year} {2017})}\BibitemShut {NoStop}%
\bibitem [{\citenamefont {Goswami}\ \emph {et~al.}(2018)\citenamefont
  {Goswami}, \citenamefont {Giarmatzi}, \citenamefont {Kewming}, \citenamefont
  {Costa}, \citenamefont {Branciard}, \citenamefont {Romero},\ and\
  \citenamefont {White}}]{Goswami}%
  \BibitemOpen
  \bibfield  {author} {\bibinfo {author} {\bibfnamefont {K.}~\bibnamefont
  {Goswami}}, \bibinfo {author} {\bibfnamefont {C.}~\bibnamefont {Giarmatzi}},
  \bibinfo {author} {\bibfnamefont {M.}~\bibnamefont {Kewming}}, \bibinfo
  {author} {\bibfnamefont {F.}~\bibnamefont {Costa}}, \bibinfo {author}
  {\bibfnamefont {C.}~\bibnamefont {Branciard}}, \bibinfo {author}
  {\bibfnamefont {J.}~\bibnamefont {Romero}},\ and\ \bibinfo {author}
  {\bibfnamefont {A.~G.}\ \bibnamefont {White}},\ }\bibfield  {title} {\bibinfo
  {title} {Indefinite causal order in a quantum switch},\ }\href
  {https://doi.org/10.1103/PhysRevLett.121.090503} {\bibfield  {journal}
  {\bibinfo  {journal} {Phys. Rev. Lett.}\ }\textbf {\bibinfo {volume} {121}},\
  \bibinfo {pages} {090503} (\bibinfo {year} {2018})}\BibitemShut {NoStop}%
\bibitem [{\citenamefont {Taddei}\ \emph {et~al.}(2021)\citenamefont {Taddei},
  \citenamefont {Cari\~ne}, \citenamefont {Mart\'{\i}nez}, \citenamefont
  {Garc\'{\i}a}, \citenamefont {Guerrero}, \citenamefont {Abbott},
  \citenamefont {Ara\'ujo}, \citenamefont {Branciard}, \citenamefont {G\'omez},
  \citenamefont {Walborn}, \citenamefont {Aolita},\ and\ \citenamefont
  {Lima}}]{Taddei}%
  \BibitemOpen
  \bibfield  {author} {\bibinfo {author} {\bibfnamefont {M.~M.}\ \bibnamefont
  {Taddei}}, \bibinfo {author} {\bibfnamefont {J.}~\bibnamefont {Cari\~ne}},
  \bibinfo {author} {\bibfnamefont {D.}~\bibnamefont {Mart\'{\i}nez}}, \bibinfo
  {author} {\bibfnamefont {T.}~\bibnamefont {Garc\'{\i}a}}, \bibinfo {author}
  {\bibfnamefont {N.}~\bibnamefont {Guerrero}}, \bibinfo {author}
  {\bibfnamefont {A.~A.}\ \bibnamefont {Abbott}}, \bibinfo {author}
  {\bibfnamefont {M.}~\bibnamefont {Ara\'ujo}}, \bibinfo {author}
  {\bibfnamefont {C.}~\bibnamefont {Branciard}}, \bibinfo {author}
  {\bibfnamefont {E.~S.}\ \bibnamefont {G\'omez}}, \bibinfo {author}
  {\bibfnamefont {S.~P.}\ \bibnamefont {Walborn}}, \bibinfo {author}
  {\bibfnamefont {L.}~\bibnamefont {Aolita}},\ and\ \bibinfo {author}
  {\bibfnamefont {G.}~\bibnamefont {Lima}},\ }\bibfield  {title} {\bibinfo
  {title} {Computational advantage from the quantum superposition of multiple
  temporal orders of photonic gates},\ }\href
  {https://doi.org/10.1103/PRXQuantum.2.010320} {\bibfield  {journal} {\bibinfo
   {journal} {PRX Quantum}\ }\textbf {\bibinfo {volume} {2}},\ \bibinfo {pages}
  {010320} (\bibinfo {year} {2021})}\BibitemShut {NoStop}%
\bibitem [{\citenamefont {Ara\'ujo}\ \emph {et~al.}(2014)\citenamefont
  {Ara\'ujo}, \citenamefont {Costa},\ and\ \citenamefont {Brukner}}]{Araujo}%
  \BibitemOpen
  \bibfield  {author} {\bibinfo {author} {\bibfnamefont {M.}~\bibnamefont
  {Ara\'ujo}}, \bibinfo {author} {\bibfnamefont {F.}~\bibnamefont {Costa}},\
  and\ \bibinfo {author} {\bibfnamefont {{\v C}.}~\bibnamefont {Brukner}},\
  }\bibfield  {title} {\bibinfo {title} {Computational advantage from
  quantum-controlled ordering of gates},\ }\href
  {https://doi.org/10.1103/PhysRevLett.113.250402} {\bibfield  {journal}
  {\bibinfo  {journal} {Phys. Rev. Lett.}\ }\textbf {\bibinfo {volume} {113}},\
  \bibinfo {pages} {250402} (\bibinfo {year} {2014})}\BibitemShut {NoStop}%
\bibitem [{\citenamefont {Gu\'erin}\ \emph {et~al.}(2016)\citenamefont
  {Gu\'erin}, \citenamefont {Feix}, \citenamefont {Ara\'ujo},\ and\
  \citenamefont {Brukner}}]{Guerin}%
  \BibitemOpen
  \bibfield  {author} {\bibinfo {author} {\bibfnamefont {P.~A.}\ \bibnamefont
  {Gu\'erin}}, \bibinfo {author} {\bibfnamefont {A.}~\bibnamefont {Feix}},
  \bibinfo {author} {\bibfnamefont {M.}~\bibnamefont {Ara\'ujo}},\ and\
  \bibinfo {author} {\bibfnamefont {{\v C}.}~\bibnamefont {Brukner}},\
  }\bibfield  {title} {\bibinfo {title} {Exponential communication complexity
  advantage from quantum superposition of the direction of communication},\
  }\href {https://doi.org/10.1103/PhysRevLett.117.100502} {\bibfield  {journal}
  {\bibinfo  {journal} {Phys. Rev. Lett.}\ }\textbf {\bibinfo {volume} {117}},\
  \bibinfo {pages} {100502} (\bibinfo {year} {2016})}\BibitemShut {NoStop}%
\bibitem [{\citenamefont {Wei}\ \emph {et~al.}(2019)\citenamefont {Wei},
  \citenamefont {Tischler}, \citenamefont {Zhao}, \citenamefont {Li},
  \citenamefont {Arrazola}, \citenamefont {Liu}, \citenamefont {Zhang},
  \citenamefont {Li}, \citenamefont {You}, \citenamefont {Wang}, \citenamefont
  {Chen}, \citenamefont {Sanders}, \citenamefont {Zhang}, \citenamefont
  {Pryde}, \citenamefont {Xu},\ and\ \citenamefont {Pan}}]{Wei}%
  \BibitemOpen
  \bibfield  {author} {\bibinfo {author} {\bibfnamefont {K.}~\bibnamefont
  {Wei}}, \bibinfo {author} {\bibfnamefont {N.}~\bibnamefont {Tischler}},
  \bibinfo {author} {\bibfnamefont {S.-R.}\ \bibnamefont {Zhao}}, \bibinfo
  {author} {\bibfnamefont {Y.-H.}\ \bibnamefont {Li}}, \bibinfo {author}
  {\bibfnamefont {J.~M.}\ \bibnamefont {Arrazola}}, \bibinfo {author}
  {\bibfnamefont {Y.}~\bibnamefont {Liu}}, \bibinfo {author} {\bibfnamefont
  {W.}~\bibnamefont {Zhang}}, \bibinfo {author} {\bibfnamefont
  {H.}~\bibnamefont {Li}}, \bibinfo {author} {\bibfnamefont {L.}~\bibnamefont
  {You}}, \bibinfo {author} {\bibfnamefont {Z.}~\bibnamefont {Wang}}, \bibinfo
  {author} {\bibfnamefont {Y.-A.}\ \bibnamefont {Chen}}, \bibinfo {author}
  {\bibfnamefont {B.~C.}\ \bibnamefont {Sanders}}, \bibinfo {author}
  {\bibfnamefont {Q.}~\bibnamefont {Zhang}}, \bibinfo {author} {\bibfnamefont
  {G.~J.}\ \bibnamefont {Pryde}}, \bibinfo {author} {\bibfnamefont
  {F.}~\bibnamefont {Xu}},\ and\ \bibinfo {author} {\bibfnamefont {J.-W.}\
  \bibnamefont {Pan}},\ }\bibfield  {title} {\bibinfo {title} {Experimental
  quantum switching for exponentially superior quantum communication
  complexity},\ }\href {https://doi.org/10.1103/PhysRevLett.122.120504}
  {\bibfield  {journal} {\bibinfo  {journal} {Phys. Rev. Lett.}\ }\textbf
  {\bibinfo {volume} {122}},\ \bibinfo {pages} {120504} (\bibinfo {year}
  {2019})}\BibitemShut {NoStop}%
\bibitem [{\citenamefont {Chiribella}(2012)}]{ChirDisc}%
  \BibitemOpen
  \bibfield  {author} {\bibinfo {author} {\bibfnamefont {G.}~\bibnamefont
  {Chiribella}},\ }\bibfield  {title} {\bibinfo {title} {Perfect discrimination
  of no-signalling channels via quantum superposition of causal structures},\
  }\href {https://doi.org/10.1103/PhysRevA.86.040301} {\bibfield  {journal}
  {\bibinfo  {journal} {Phys. Rev. A}\ }\textbf {\bibinfo {volume} {86}},\
  \bibinfo {pages} {040301(R)} (\bibinfo {year} {2012})}\BibitemShut {NoStop}%
\bibitem [{\citenamefont {Goswami}\ \emph {et~al.}(2020)\citenamefont
  {Goswami}, \citenamefont {Cao}, \citenamefont {Paz-Silva}, \citenamefont
  {Romero},\ and\ \citenamefont {White}}]{Goswami2020}%
  \BibitemOpen
  \bibfield  {author} {\bibinfo {author} {\bibfnamefont {K.}~\bibnamefont
  {Goswami}}, \bibinfo {author} {\bibfnamefont {Y.}~\bibnamefont {Cao}},
  \bibinfo {author} {\bibfnamefont {G.~A.}\ \bibnamefont {Paz-Silva}}, \bibinfo
  {author} {\bibfnamefont {J.}~\bibnamefont {Romero}},\ and\ \bibinfo {author}
  {\bibfnamefont {A.~G.}\ \bibnamefont {White}},\ }\bibfield  {title} {\bibinfo
  {title} {Increasing communication capacity via superposition of order},\
  }\href {https://doi.org/10.1103/PhysRevResearch.2.033292} {\bibfield
  {journal} {\bibinfo  {journal} {Phys. Rev. Res.}\ }\textbf {\bibinfo {volume}
  {2}},\ \bibinfo {pages} {033292} (\bibinfo {year} {2020})}\BibitemShut
  {NoStop}%
\bibitem [{\citenamefont {Guo}\ \emph {et~al.}(2020)\citenamefont {Guo},
  \citenamefont {Hu}, \citenamefont {Hou}, \citenamefont {Cao}, \citenamefont
  {Cui}, \citenamefont {Liu}, \citenamefont {Huang}, \citenamefont {Li},
  \citenamefont {Guo},\ and\ \citenamefont {Chiribella}}]{Guo2020}%
  \BibitemOpen
  \bibfield  {author} {\bibinfo {author} {\bibfnamefont {Y.}~\bibnamefont
  {Guo}}, \bibinfo {author} {\bibfnamefont {X.-M.}\ \bibnamefont {Hu}},
  \bibinfo {author} {\bibfnamefont {Z.-B.}\ \bibnamefont {Hou}}, \bibinfo
  {author} {\bibfnamefont {H.}~\bibnamefont {Cao}}, \bibinfo {author}
  {\bibfnamefont {J.-M.}\ \bibnamefont {Cui}}, \bibinfo {author} {\bibfnamefont
  {B.-H.}\ \bibnamefont {Liu}}, \bibinfo {author} {\bibfnamefont {Y.-F.}\
  \bibnamefont {Huang}}, \bibinfo {author} {\bibfnamefont {C.-F.}\ \bibnamefont
  {Li}}, \bibinfo {author} {\bibfnamefont {G.-C.}\ \bibnamefont {Guo}},\ and\
  \bibinfo {author} {\bibfnamefont {G.}~\bibnamefont {Chiribella}},\ }\bibfield
   {title} {\bibinfo {title} {Experimental transmission of quantum information
  using a superposition of causal orders},\ }\href
  {https://doi.org/10.1103/PhysRevLett.124.030502} {\bibfield  {journal}
  {\bibinfo  {journal} {Phys. Rev. Lett.}\ }\textbf {\bibinfo {volume} {124}},\
  \bibinfo {pages} {030502} (\bibinfo {year} {2020})}\BibitemShut {NoStop}%
\bibitem [{\citenamefont {Felce}\ and\ \citenamefont
  {Vedral}(2020)}]{Refrigerator_Vedral}%
  \BibitemOpen
  \bibfield  {author} {\bibinfo {author} {\bibfnamefont {D.}~\bibnamefont
  {Felce}}\ and\ \bibinfo {author} {\bibfnamefont {V.}~\bibnamefont {Vedral}},\
  }\bibfield  {title} {\bibinfo {title} {Quantum refrigeration with indefinite
  causal order},\ }\href {https://doi.org/10.1103/PhysRevLett.125.070603}
  {\bibfield  {journal} {\bibinfo  {journal} {Phys. Rev. Lett.}\ }\textbf
  {\bibinfo {volume} {125}},\ \bibinfo {pages} {070603} (\bibinfo {year}
  {2020})}\BibitemShut {NoStop}%
\bibitem [{\citenamefont {Nie}\ \emph {et~al.}()\citenamefont {Nie},
  \citenamefont {Zhu}, \citenamefont {Xi}, \citenamefont {Long}, \citenamefont
  {Lin}, \citenamefont {Tian}, \citenamefont {Qiu}, \citenamefont {Yang},
  \citenamefont {Dong}, \citenamefont {Li}, \citenamefont {Xin},\ and\
  \citenamefont {Lu}}]{RefrigExp}%
  \BibitemOpen
  \bibfield  {author} {\bibinfo {author} {\bibfnamefont {X.}~\bibnamefont
  {Nie}}, \bibinfo {author} {\bibfnamefont {X.}~\bibnamefont {Zhu}}, \bibinfo
  {author} {\bibfnamefont {C.}~\bibnamefont {Xi}}, \bibinfo {author}
  {\bibfnamefont {X.}~\bibnamefont {Long}}, \bibinfo {author} {\bibfnamefont
  {Z.}~\bibnamefont {Lin}}, \bibinfo {author} {\bibfnamefont {Y.}~\bibnamefont
  {Tian}}, \bibinfo {author} {\bibfnamefont {C.}~\bibnamefont {Qiu}}, \bibinfo
  {author} {\bibfnamefont {X.}~\bibnamefont {Yang}}, \bibinfo {author}
  {\bibfnamefont {Y.}~\bibnamefont {Dong}}, \bibinfo {author} {\bibfnamefont
  {J.}~\bibnamefont {Li}}, \bibinfo {author} {\bibfnamefont {T.}~\bibnamefont
  {Xin}},\ and\ \bibinfo {author} {\bibfnamefont {D.}~\bibnamefont {Lu}},\
  }\href@noop {} {\bibinfo {title} {Experimental realization of a quantum
  refrigerator driven by indefinite causal orders}},\ \Eprint
  {https://arxiv.org/abs/2011.12580} {arXiv:2011.12580} \BibitemShut {NoStop}%
\bibitem [{\citenamefont {Zhao}\ \emph {et~al.}(2020)\citenamefont {Zhao},
  \citenamefont {Yang},\ and\ \citenamefont {Chiribella}}]{Zhao2020}%
  \BibitemOpen
  \bibfield  {author} {\bibinfo {author} {\bibfnamefont {X.}~\bibnamefont
  {Zhao}}, \bibinfo {author} {\bibfnamefont {Y.}~\bibnamefont {Yang}},\ and\
  \bibinfo {author} {\bibfnamefont {G.}~\bibnamefont {Chiribella}},\ }\bibfield
   {title} {\bibinfo {title} {Quantum metrology with indefinite causal order},\
  }\href {https://doi.org/10.1103/PhysRevLett.124.190503} {\bibfield  {journal}
  {\bibinfo  {journal} {Phys. Rev. Lett.}\ }\textbf {\bibinfo {volume} {124}},\
  \bibinfo {pages} {190503} (\bibinfo {year} {2020})}\BibitemShut {NoStop}%
\bibitem [{\citenamefont {Chiribella}\ \emph {et~al.}(2008)\citenamefont
  {Chiribella}, \citenamefont {D’Ariano},\ and\ \citenamefont
  {Perinotti}}]{ChirSupermap}%
  \BibitemOpen
  \bibfield  {author} {\bibinfo {author} {\bibfnamefont {G.}~\bibnamefont
  {Chiribella}}, \bibinfo {author} {\bibfnamefont {G.~M.}\ \bibnamefont
  {D’Ariano}},\ and\ \bibinfo {author} {\bibfnamefont {P.}~\bibnamefont
  {Perinotti}},\ }\bibfield  {title} {\bibinfo {title} {Transforming quantum
  operations: Quantum supermaps},\ }\href
  {https://doi.org/10.1209/0295-5075/83/30004} {\bibfield  {journal} {\bibinfo
  {journal} {Europhys. Lett.}\ }\textbf {\bibinfo {volume} {83}},\ \bibinfo
  {pages} {30004} (\bibinfo {year} {2008})}\BibitemShut {NoStop}%
\bibitem [{\citenamefont {Oreshkov}\ \emph {et~al.}(2012)\citenamefont
  {Oreshkov}, \citenamefont {Costa},\ and\ \citenamefont {Brukner}}]{Oreshkov}%
  \BibitemOpen
  \bibfield  {author} {\bibinfo {author} {\bibfnamefont {O.}~\bibnamefont
  {Oreshkov}}, \bibinfo {author} {\bibfnamefont {F.}~\bibnamefont {Costa}},\
  and\ \bibinfo {author} {\bibfnamefont {{\v C}.}~\bibnamefont {Brukner}},\
  }\bibfield  {title} {\bibinfo {title} {Quantum correlations with no causal
  order},\ }\href {https://doi.org/10.1038/ncomms2076} {\bibfield  {journal}
  {\bibinfo  {journal} {Nat. Commun.}\ }\textbf {\bibinfo {volume} {3}},\
  \bibinfo {pages} {1092} (\bibinfo {year} {2012})}\BibitemShut {NoStop}%
\bibitem [{\citenamefont {Hardy}(2007)}]{Hardy}%
  \BibitemOpen
  \bibfield  {author} {\bibinfo {author} {\bibfnamefont {L.}~\bibnamefont
  {Hardy}},\ }\bibfield  {title} {\bibinfo {title} {Towards quantum gravity: A
  framework for probabilistic theories with non-fixed causal structure},\
  }\href {https://doi.org/10.1088/1751-8113/40/12/s12} {\bibfield  {journal}
  {\bibinfo  {journal} {J. Phys. A: Math. Theor.}\ }\textbf {\bibinfo {volume}
  {40}},\ \bibinfo {pages} {3081–3099} (\bibinfo {year} {2007})}\BibitemShut
  {NoStop}%
\bibitem [{\citenamefont {Ara{\'{u}}jo}\ \emph {et~al.}(2015)\citenamefont
  {Ara{\'{u}}jo}, \citenamefont {Branciard}, \citenamefont {Costa},
  \citenamefont {Feix}, \citenamefont {Giarmatzi},\ and\ \citenamefont
  {Brukner}}]{Araujo2015}%
  \BibitemOpen
  \bibfield  {author} {\bibinfo {author} {\bibfnamefont {M.}~\bibnamefont
  {Ara{\'{u}}jo}}, \bibinfo {author} {\bibfnamefont {C.}~\bibnamefont
  {Branciard}}, \bibinfo {author} {\bibfnamefont {F.}~\bibnamefont {Costa}},
  \bibinfo {author} {\bibfnamefont {A.}~\bibnamefont {Feix}}, \bibinfo {author}
  {\bibfnamefont {C.}~\bibnamefont {Giarmatzi}},\ and\ \bibinfo {author}
  {\bibfnamefont {{\v{C}}.}~\bibnamefont {Brukner}},\ }\bibfield  {title}
  {\bibinfo {title} {Witnessing causal nonseparability},\ }\href
  {https://doi.org/10.1088/1367-2630/17/10/102001} {\bibfield  {journal}
  {\bibinfo  {journal} {New Journal of Physics}\ }\textbf {\bibinfo {volume}
  {17}},\ \bibinfo {pages} {102001} (\bibinfo {year} {2015})}\BibitemShut
  {NoStop}%
\bibitem [{\citenamefont {Zych}\ \emph {et~al.}(2019)\citenamefont {Zych},
  \citenamefont {Costa}, \citenamefont {Pikovski},\ and\ \citenamefont
  {Brukner}}]{tbell}%
  \BibitemOpen
  \bibfield  {author} {\bibinfo {author} {\bibfnamefont {M.}~\bibnamefont
  {Zych}}, \bibinfo {author} {\bibfnamefont {F.}~\bibnamefont {Costa}},
  \bibinfo {author} {\bibfnamefont {I.}~\bibnamefont {Pikovski}},\ and\
  \bibinfo {author} {\bibfnamefont {{\v C}.}~\bibnamefont {Brukner}},\
  }\bibfield  {title} {\bibinfo {title} {Bell’s theorem for temporal order},\
  }\href {https://doi.org/10.1038/s41467-019-11579-x} {\bibfield  {journal}
  {\bibinfo  {journal} {Nat. Commun.}\ }\textbf {\bibinfo {volume} {10}},\
  \bibinfo {pages} {3772} (\bibinfo {year} {2019})}\BibitemShut {NoStop}%
\bibitem [{\citenamefont {Rubino}\ \emph {et~al.}()\citenamefont {Rubino},
  \citenamefont {Rozema}, \citenamefont {Massa}, \citenamefont {Ara\'{u}jo},
  \citenamefont {Magdalena}, \citenamefont {\v{C}aslav Brukner},\ and\
  \citenamefont {Walther}}]{RubinoAgain}%
  \BibitemOpen
  \bibfield  {author} {\bibinfo {author} {\bibfnamefont {G.}~\bibnamefont
  {Rubino}}, \bibinfo {author} {\bibfnamefont {L.~A.}\ \bibnamefont {Rozema}},
  \bibinfo {author} {\bibfnamefont {F.}~\bibnamefont {Massa}}, \bibinfo
  {author} {\bibfnamefont {M.}~\bibnamefont {Ara\'{u}jo}}, \bibinfo {author}
  {\bibnamefont {Magdalena}}, \bibinfo {author} {\bibnamefont {\v{C}aslav
  Brukner}},\ and\ \bibinfo {author} {\bibfnamefont {P.}~\bibnamefont
  {Walther}},\ }\href@noop {} {\bibinfo {title} {Experimental entanglement of
  temporal orders}},\ \Eprint {https://arxiv.org/abs/1712.06884}
  {arXiv:1712.06884} \BibitemShut {NoStop}%
\bibitem [{\citenamefont {Ford}(1982)}]{FORD1982238}%
  \BibitemOpen
  \bibfield  {author} {\bibinfo {author} {\bibfnamefont {L.}~\bibnamefont
  {Ford}},\ }\bibfield  {title} {\bibinfo {title} {Gravitational radiation by
  quantum systems},\ }\href
  {https://doi.org/https://doi.org/10.1016/0003-4916(82)90115-4} {\bibfield
  {journal} {\bibinfo  {journal} {Annals of Physics}\ }\textbf {\bibinfo
  {volume} {144}},\ \bibinfo {pages} {238 } (\bibinfo {year}
  {1982})}\BibitemShut {NoStop}%
\bibitem [{\citenamefont {Anastopoulos}\ and\ \citenamefont
  {Hu}(2015)}]{Anastopoulos_2015}%
  \BibitemOpen
  \bibfield  {author} {\bibinfo {author} {\bibfnamefont {C.}~\bibnamefont
  {Anastopoulos}}\ and\ \bibinfo {author} {\bibfnamefont {B.~L.}\ \bibnamefont
  {Hu}},\ }\bibfield  {title} {\bibinfo {title} {Probing a gravitational cat
  state},\ }\href {https://doi.org/10.1088/0264-9381/32/16/165022} {\bibfield
  {journal} {\bibinfo  {journal} {Classic. Quantum Grav.}\ }\textbf {\bibinfo
  {volume} {32}},\ \bibinfo {pages} {165022} (\bibinfo {year}
  {2015})}\BibitemShut {NoStop}%
\bibitem [{\citenamefont {Bose}\ \emph {et~al.}(2017)\citenamefont {Bose},
  \citenamefont {Mazumdar}, \citenamefont {Morley}, \citenamefont {Ulbricht},
  \citenamefont {Toro\ifmmode~\check{s}\else \v{s}\fi{}}, \citenamefont
  {Paternostro}, \citenamefont {Geraci}, \citenamefont {Barker}, \citenamefont
  {Kim},\ and\ \citenamefont {Milburn}}]{Bose}%
  \BibitemOpen
  \bibfield  {author} {\bibinfo {author} {\bibfnamefont {S.}~\bibnamefont
  {Bose}}, \bibinfo {author} {\bibfnamefont {A.}~\bibnamefont {Mazumdar}},
  \bibinfo {author} {\bibfnamefont {G.~W.}\ \bibnamefont {Morley}}, \bibinfo
  {author} {\bibfnamefont {H.}~\bibnamefont {Ulbricht}}, \bibinfo {author}
  {\bibfnamefont {M.}~\bibnamefont {Toro\ifmmode~\check{s}\else \v{s}\fi{}}},
  \bibinfo {author} {\bibfnamefont {M.}~\bibnamefont {Paternostro}}, \bibinfo
  {author} {\bibfnamefont {A.~A.}\ \bibnamefont {Geraci}}, \bibinfo {author}
  {\bibfnamefont {P.~F.}\ \bibnamefont {Barker}}, \bibinfo {author}
  {\bibfnamefont {M.~S.}\ \bibnamefont {Kim}},\ and\ \bibinfo {author}
  {\bibfnamefont {G.}~\bibnamefont {Milburn}},\ }\bibfield  {title} {\bibinfo
  {title} {Spin entanglement witness for quantum gravity},\ }\href
  {https://doi.org/10.1103/PhysRevLett.119.240401} {\bibfield  {journal}
  {\bibinfo  {journal} {Phys. Rev. Lett.}\ }\textbf {\bibinfo {volume} {119}},\
  \bibinfo {pages} {240401} (\bibinfo {year} {2017})}\BibitemShut {NoStop}%
\bibitem [{\citenamefont {Marletto}\ and\ \citenamefont
  {Vedral}(2017)}]{Marletto}%
  \BibitemOpen
  \bibfield  {author} {\bibinfo {author} {\bibfnamefont {C.}~\bibnamefont
  {Marletto}}\ and\ \bibinfo {author} {\bibfnamefont {V.}~\bibnamefont
  {Vedral}},\ }\bibfield  {title} {\bibinfo {title} {Gravitationally induced
  entanglement between two massive particles is sufficient evidence of quantum
  effects in gravity},\ }\href {https://doi.org/10.1103/PhysRevLett.119.240402}
  {\bibfield  {journal} {\bibinfo  {journal} {Phys. Rev. Lett.}\ }\textbf
  {\bibinfo {volume} {119}},\ \bibinfo {pages} {240402} (\bibinfo {year}
  {2017})}\BibitemShut {NoStop}%
\bibitem [{\citenamefont {Belenchia}\ \emph {et~al.}(2018)\citenamefont
  {Belenchia}, \citenamefont {Wald}, \citenamefont {Giacomini}, \citenamefont
  {Castro-Ruiz}, \citenamefont {Brukner},\ and\ \citenamefont
  {Aspelmeyer}}]{Belechia_Wald_2018}%
  \BibitemOpen
  \bibfield  {author} {\bibinfo {author} {\bibfnamefont {A.}~\bibnamefont
  {Belenchia}}, \bibinfo {author} {\bibfnamefont {R.~M.}\ \bibnamefont {Wald}},
  \bibinfo {author} {\bibfnamefont {F.}~\bibnamefont {Giacomini}}, \bibinfo
  {author} {\bibfnamefont {E.}~\bibnamefont {Castro-Ruiz}}, \bibinfo {author}
  {\bibfnamefont {{\v C}.}~\bibnamefont {Brukner}},\ and\ \bibinfo {author}
  {\bibfnamefont {M.}~\bibnamefont {Aspelmeyer}},\ }\bibfield  {title}
  {\bibinfo {title} {Quantum superposition of massive objects and the
  quantization of gravity},\ }\href
  {https://doi.org/10.1103/PhysRevD.98.126009} {\bibfield  {journal} {\bibinfo
  {journal} {Phys. Rev. D}\ }\textbf {\bibinfo {volume} {98}},\ \bibinfo
  {pages} {126009} (\bibinfo {year} {2018})}\BibitemShut {NoStop}%
\bibitem [{\citenamefont {Christodoulou}\ and\ \citenamefont
  {Rovelli}(2019)}]{Rovelli}%
  \BibitemOpen
  \bibfield  {author} {\bibinfo {author} {\bibfnamefont {M.}~\bibnamefont
  {Christodoulou}}\ and\ \bibinfo {author} {\bibfnamefont {C.}~\bibnamefont
  {Rovelli}},\ }\bibfield  {title} {\bibinfo {title} {On the possibility of
  laboratory evidence for quantum superposition of geometries},\ }\href
  {https://doi.org/https://doi.org/10.1016/j.physletb.2019.03.015} {\bibfield
  {journal} {\bibinfo  {journal} {Physics Letters B}\ }\textbf {\bibinfo
  {volume} {792}},\ \bibinfo {pages} {64 } (\bibinfo {year}
  {2019})}\BibitemShut {NoStop}%
\bibitem [{\citenamefont {Howl}\ \emph {et~al.}(2021)\citenamefont {Howl},
  \citenamefont {Vedral}, \citenamefont {Naik}, \citenamefont {Christodoulou},
  \citenamefont {Rovelli},\ and\ \citenamefont {Iyer}}]{Howl2020}%
  \BibitemOpen
  \bibfield  {author} {\bibinfo {author} {\bibfnamefont {R.}~\bibnamefont
  {Howl}}, \bibinfo {author} {\bibfnamefont {V.}~\bibnamefont {Vedral}},
  \bibinfo {author} {\bibfnamefont {D.}~\bibnamefont {Naik}}, \bibinfo {author}
  {\bibfnamefont {M.}~\bibnamefont {Christodoulou}}, \bibinfo {author}
  {\bibfnamefont {C.}~\bibnamefont {Rovelli}},\ and\ \bibinfo {author}
  {\bibfnamefont {A.}~\bibnamefont {Iyer}},\ }\bibfield  {title} {\bibinfo
  {title} {Non-gaussianity as a signature of a quantum theory of gravity},\
  }\href {https://doi.org/10.1103/PRXQuantum.2.010325} {\bibfield  {journal}
  {\bibinfo  {journal} {PRX Quantum}\ }\textbf {\bibinfo {volume} {2}},\
  \bibinfo {pages} {010325} (\bibinfo {year} {2021})}\BibitemShut {NoStop}%
\bibitem [{\citenamefont {Paunković}\ and\ \citenamefont
  {Vojinović}(2020)}]{Nikola}%
  \BibitemOpen
  \bibfield  {author} {\bibinfo {author} {\bibfnamefont {N.}~\bibnamefont
  {Paunković}}\ and\ \bibinfo {author} {\bibfnamefont {M.}~\bibnamefont
  {Vojinović}},\ }\bibfield  {title} {\bibinfo {title} {Causal orders, quantum
  circuits and spacetime: Distinguishing between definite and superposed causal
  orders},\ }\href {https://doi.org/10.22331/q-2020-05-28-275} {\bibfield
  {journal} {\bibinfo  {journal} {Quantum}\ }\textbf {\bibinfo {volume} {4}},\
  \bibinfo {pages} {275} (\bibinfo {year} {2020})}\BibitemShut {NoStop}%
\bibitem [{\citenamefont {Dimić}\ \emph {et~al.}(2020)\citenamefont {Dimić},
  \citenamefont {Milivojević}, \citenamefont {Go{\v c}anin}, \citenamefont
  {M\'oller},\ and\ \citenamefont {Brukner}}]{Rindler}%
  \BibitemOpen
  \bibfield  {author} {\bibinfo {author} {\bibfnamefont {A.}~\bibnamefont
  {Dimić}}, \bibinfo {author} {\bibfnamefont {M.}~\bibnamefont
  {Milivojević}}, \bibinfo {author} {\bibfnamefont {D.}~\bibnamefont {Go{\v
  c}anin}}, \bibinfo {author} {\bibfnamefont {N.~S.}\ \bibnamefont
  {M\'oller}},\ and\ \bibinfo {author} {\bibfnamefont {{\v C}.}~\bibnamefont
  {Brukner}},\ }\bibfield  {title} {\bibinfo {title} {Simulating indefinite
  causal order with rindler observers},\ }\href
  {https://doi.org/10.3389/fphy.2020.525333} {\bibfield  {journal} {\bibinfo
  {journal} {Front. Phys.}\ }\textbf {\bibinfo {volume} {8}},\ \bibinfo {pages}
  {470} (\bibinfo {year} {2020})}\BibitemShut {NoStop}%
\bibitem [{\citenamefont {Zych}\ \emph {et~al.}()\citenamefont {Zych},
  \citenamefont {Costa},\ and\ \citenamefont {Ralph}}]{ZychRelQuantSup}%
  \BibitemOpen
  \bibfield  {author} {\bibinfo {author} {\bibfnamefont {M.}~\bibnamefont
  {Zych}}, \bibinfo {author} {\bibfnamefont {F.}~\bibnamefont {Costa}},\ and\
  \bibinfo {author} {\bibfnamefont {T.~C.}\ \bibnamefont {Ralph}},\ }\href@noop
  {} {\bibinfo {title} {Relativity of quantum superpositions}},\ \Eprint
  {https://arxiv.org/abs/1809.04999} {arXiv:1809.04999} \BibitemShut {NoStop}%
\bibitem [{\citenamefont {Zych}(2017)}]{Zych}%
  \BibitemOpen
  \bibfield  {author} {\bibinfo {author} {\bibfnamefont {M.}~\bibnamefont
  {Zych}},\ }\href {https://doi.org/10.1007/978-3-319-53192-2} {\emph {\bibinfo
  {title} {Quantum Systems under Gravitational Time Dilation}}}\ (\bibinfo
  {publisher} {Springer, Cham, Switzerland},\ \bibinfo {year}
  {2017})\BibitemShut {NoStop}%
\bibitem [{\citenamefont {Zych}\ \emph {et~al.}(2011)\citenamefont {Zych},
  \citenamefont {Costa}, \citenamefont {Pikovski},\ and\ \citenamefont
  {Brukner}}]{Zych2011}%
  \BibitemOpen
  \bibfield  {author} {\bibinfo {author} {\bibfnamefont {M.}~\bibnamefont
  {Zych}}, \bibinfo {author} {\bibfnamefont {F.}~\bibnamefont {Costa}},
  \bibinfo {author} {\bibfnamefont {I.}~\bibnamefont {Pikovski}},\ and\
  \bibinfo {author} {\bibfnamefont {{\v C}.}~\bibnamefont {Brukner}},\
  }\bibfield  {title} {\bibinfo {title} {Quantum interferometric visibility as
  a witness of general relativistic proper time},\ }\href
  {https://doi.org/10.1038/ncomms1498} {\bibfield  {journal} {\bibinfo
  {journal} {Nat. Commun.}\ }\textbf {\bibinfo {volume} {2}},\ \bibinfo {pages}
  {505} (\bibinfo {year} {2011})}\BibitemShut {NoStop}%
\bibitem [{\citenamefont {Zych}\ \emph {et~al.}(2012)\citenamefont {Zych},
  \citenamefont {Costa}, \citenamefont {Pikovski}, \citenamefont {Ralph},\ and\
  \citenamefont {Brukner}}]{Zych_2012}%
  \BibitemOpen
  \bibfield  {author} {\bibinfo {author} {\bibfnamefont {M.}~\bibnamefont
  {Zych}}, \bibinfo {author} {\bibfnamefont {F.}~\bibnamefont {Costa}},
  \bibinfo {author} {\bibfnamefont {I.}~\bibnamefont {Pikovski}}, \bibinfo
  {author} {\bibfnamefont {T.~C.}\ \bibnamefont {Ralph}},\ and\ \bibinfo
  {author} {\bibfnamefont {{\v{C}}.}~\bibnamefont {Brukner}},\ }\bibfield
  {title} {\bibinfo {title} {General relativistic effects in quantum
  interference of photons},\ }\href
  {https://doi.org/10.1088/0264-9381/29/22/224010} {\bibfield  {journal}
  {\bibinfo  {journal} {Classic. Quantum Grav.}\ }\textbf {\bibinfo {volume}
  {29}},\ \bibinfo {pages} {224010} (\bibinfo {year} {2012})}\BibitemShut
  {NoStop}%
\bibitem [{\citenamefont {Pikovski}\ \emph {et~al.}(2015)\citenamefont
  {Pikovski}, \citenamefont {Zych}, \citenamefont {Costa},\ and\ \citenamefont
  {Brukner}}]{Pikovski2015}%
  \BibitemOpen
  \bibfield  {author} {\bibinfo {author} {\bibfnamefont {I.}~\bibnamefont
  {Pikovski}}, \bibinfo {author} {\bibfnamefont {M.}~\bibnamefont {Zych}},
  \bibinfo {author} {\bibfnamefont {F.}~\bibnamefont {Costa}},\ and\ \bibinfo
  {author} {\bibfnamefont {{\v C}.}~\bibnamefont {Brukner}},\ }\bibfield
  {title} {\bibinfo {title} {Universal decoherence due to gravitational time
  dilation},\ }\href {https://doi.org/10.1038/nphys3366} {\bibfield  {journal}
  {\bibinfo  {journal} {Nature Physics}\ }\textbf {\bibinfo {volume} {11}},\
  \bibinfo {pages} {668} (\bibinfo {year} {2015})}\BibitemShut {NoStop}%
\bibitem [{\citenamefont {Margalit}\ \emph {et~al.}(2015)\citenamefont
  {Margalit}, \citenamefont {Zhou}, \citenamefont {Machluf}, \citenamefont
  {Rohrlich}, \citenamefont {Japha},\ and\ \citenamefont {Folman}}]{Margalit}%
  \BibitemOpen
  \bibfield  {author} {\bibinfo {author} {\bibfnamefont {Y.}~\bibnamefont
  {Margalit}}, \bibinfo {author} {\bibfnamefont {Z.}~\bibnamefont {Zhou}},
  \bibinfo {author} {\bibfnamefont {S.}~\bibnamefont {Machluf}}, \bibinfo
  {author} {\bibfnamefont {D.}~\bibnamefont {Rohrlich}}, \bibinfo {author}
  {\bibfnamefont {Y.}~\bibnamefont {Japha}},\ and\ \bibinfo {author}
  {\bibfnamefont {R.}~\bibnamefont {Folman}},\ }\bibfield  {title} {\bibinfo
  {title} {A self-interfering clock as a “which path” witness},\ }\href
  {https://doi.org/10.1126/science.aac6498} {\bibfield  {journal} {\bibinfo
  {journal} {Science}\ }\textbf {\bibinfo {volume} {349}},\ \bibinfo {pages}
  {1205–1208} (\bibinfo {year} {2015})}\BibitemShut {NoStop}%
\bibitem [{\citenamefont {Lammerzahl}(1995)}]{Lammerzahl:1995zz}%
  \BibitemOpen
  \bibfield  {author} {\bibinfo {author} {\bibfnamefont {C.}~\bibnamefont
  {Lammerzahl}},\ }\bibfield  {title} {\bibinfo {title} {{A Hamilton operator
  for quantum optics in gravitational fields}},\ }\href
  {https://doi.org/10.1016/0375-9601(95)00345-4} {\bibfield  {journal}
  {\bibinfo  {journal} {Phys. Lett. A}\ }\textbf {\bibinfo {volume} {203}},\
  \bibinfo {pages} {12} (\bibinfo {year} {1995})}\BibitemShut {NoStop}%
\bibitem [{\citenamefont {Kiefer}\ and\ \citenamefont
  {Singh}(1991)}]{Kiefer1991}%
  \BibitemOpen
  \bibfield  {author} {\bibinfo {author} {\bibfnamefont {C.}~\bibnamefont
  {Kiefer}}\ and\ \bibinfo {author} {\bibfnamefont {T.~P.}\ \bibnamefont
  {Singh}},\ }\bibfield  {title} {\bibinfo {title} {Quantum gravitational
  corrections to the functional schr\"odinger equation},\ }\href
  {https://doi.org/10.1103/PhysRevD.44.1067} {\bibfield  {journal} {\bibinfo
  {journal} {Phys. Rev. D}\ }\textbf {\bibinfo {volume} {44}},\ \bibinfo
  {pages} {1067} (\bibinfo {year} {1991})}\BibitemShut {NoStop}%
\bibitem [{\citenamefont {Goswami}\ and\ \citenamefont
  {Romero}(2020)}]{ReviewExp}%
  \BibitemOpen
  \bibfield  {author} {\bibinfo {author} {\bibfnamefont {K.}~\bibnamefont
  {Goswami}}\ and\ \bibinfo {author} {\bibfnamefont {J.}~\bibnamefont
  {Romero}},\ }\bibfield  {title} {\bibinfo {title} {Experiments on quantum
  causality},\ }\href {https://doi.org/10.1116/5.0010747} {\bibfield  {journal}
  {\bibinfo  {journal} {AVS Quantum Sci.}\ }\textbf {\bibinfo {volume} {2}},\
  \bibinfo {pages} {037101} (\bibinfo {year} {2020})}\BibitemShut {NoStop}%
\bibitem [{\citenamefont {Overhauser}\ and\ \citenamefont
  {Colella}(1974)}]{Collela}%
  \BibitemOpen
  \bibfield  {author} {\bibinfo {author} {\bibfnamefont {A.~W.}\ \bibnamefont
  {Overhauser}}\ and\ \bibinfo {author} {\bibfnamefont {R.}~\bibnamefont
  {Colella}},\ }\bibfield  {title} {\bibinfo {title} {Experimental test of
  gravitationally induced quantum interference},\ }\href
  {https://doi.org/10.1103/PhysRevLett.33.1237} {\bibfield  {journal} {\bibinfo
   {journal} {Phys. Rev. Lett.}\ }\textbf {\bibinfo {volume} {33}},\ \bibinfo
  {pages} {1237} (\bibinfo {year} {1974})}\BibitemShut {NoStop}%
\bibitem [{\citenamefont {Nesvizhevsky}\ \emph {et~al.}(2002)\citenamefont
  {Nesvizhevsky}, \citenamefont {Börner}, \citenamefont {Petukhov},
  \citenamefont {Abele}, \citenamefont {Baessler}, \citenamefont {Ruess},
  \citenamefont {Stöferle}, \citenamefont {Westphal}, \citenamefont
  {Gagarski}, \citenamefont {Petrov},\ and\ \citenamefont
  {Strelkov}}]{Strelkov}%
  \BibitemOpen
  \bibfield  {author} {\bibinfo {author} {\bibfnamefont {V.}~\bibnamefont
  {Nesvizhevsky}}, \bibinfo {author} {\bibfnamefont {H.}~\bibnamefont
  {Börner}}, \bibinfo {author} {\bibfnamefont {A.}~\bibnamefont {Petukhov}},
  \bibinfo {author} {\bibfnamefont {H.}~\bibnamefont {Abele}}, \bibinfo
  {author} {\bibfnamefont {S.}~\bibnamefont {Baessler}}, \bibinfo {author}
  {\bibfnamefont {F.}~\bibnamefont {Ruess}}, \bibinfo {author} {\bibfnamefont
  {T.}~\bibnamefont {Stöferle}}, \bibinfo {author} {\bibfnamefont
  {A.}~\bibnamefont {Westphal}}, \bibinfo {author} {\bibfnamefont
  {A.}~\bibnamefont {Gagarski}}, \bibinfo {author} {\bibfnamefont
  {G.}~\bibnamefont {Petrov}},\ and\ \bibinfo {author} {\bibfnamefont
  {A.}~\bibnamefont {Strelkov}},\ }\bibfield  {title} {\bibinfo {title}
  {Quantum states of neutrons in the earth's gravitational field},\ }\href
  {https://doi.org/10.1038/415297a} {\bibfield  {journal} {\bibinfo  {journal}
  {Nature}\ }\textbf {\bibinfo {volume} {415}},\ \bibinfo {pages} {297}
  (\bibinfo {year} {2002})}\BibitemShut {NoStop}%
\bibitem [{\citenamefont {Chou}\ \emph {et~al.}(2010)\citenamefont {Chou},
  \citenamefont {Hume}, \citenamefont {Rosenband},\ and\ \citenamefont
  {Wineland}}]{Chou2010}%
  \BibitemOpen
  \bibfield  {author} {\bibinfo {author} {\bibfnamefont {C.~W.}\ \bibnamefont
  {Chou}}, \bibinfo {author} {\bibfnamefont {D.~B.}\ \bibnamefont {Hume}},
  \bibinfo {author} {\bibfnamefont {T.}~\bibnamefont {Rosenband}},\ and\
  \bibinfo {author} {\bibfnamefont {D.~J.}\ \bibnamefont {Wineland}},\
  }\bibfield  {title} {\bibinfo {title} {Optical clocks and relativity},\
  }\href {https://doi.org/10.1126/science.1192720} {\bibfield  {journal}
  {\bibinfo  {journal} {Science}\ }\textbf {\bibinfo {volume} {329}},\ \bibinfo
  {pages} {1630} (\bibinfo {year} {2010})}\BibitemShut {NoStop}%
\bibitem [{\citenamefont {Brodutch}\ \emph {et~al.}(2015)\citenamefont
  {Brodutch}, \citenamefont {Gilchrist}, \citenamefont {Guff}, \citenamefont
  {Smith},\ and\ \citenamefont {Terno}}]{Terno2015}%
  \BibitemOpen
  \bibfield  {author} {\bibinfo {author} {\bibfnamefont {A.}~\bibnamefont
  {Brodutch}}, \bibinfo {author} {\bibfnamefont {A.}~\bibnamefont {Gilchrist}},
  \bibinfo {author} {\bibfnamefont {T.}~\bibnamefont {Guff}}, \bibinfo {author}
  {\bibfnamefont {A.~R.~H.}\ \bibnamefont {Smith}},\ and\ \bibinfo {author}
  {\bibfnamefont {D.~R.}\ \bibnamefont {Terno}},\ }\bibfield  {title} {\bibinfo
  {title} {Post-newtonian gravitational effects in optical interferometry},\
  }\href {https://doi.org/10.1103/PhysRevD.91.064041} {\bibfield  {journal}
  {\bibinfo  {journal} {Phys. Rev. D}\ }\textbf {\bibinfo {volume} {91}},\
  \bibinfo {pages} {064041} (\bibinfo {year} {2015})}\BibitemShut {NoStop}%
\bibitem [{\citenamefont {Rivera-Tapia}\ \emph {et~al.}(2020)\citenamefont
  {Rivera-Tapia}, \citenamefont {Delgado},\ and\ \citenamefont
  {Rubilar}}]{Rivera_Tapia_2020}%
  \BibitemOpen
  \bibfield  {author} {\bibinfo {author} {\bibfnamefont {M.}~\bibnamefont
  {Rivera-Tapia}}, \bibinfo {author} {\bibfnamefont {A.}~\bibnamefont
  {Delgado}},\ and\ \bibinfo {author} {\bibfnamefont {G.}~\bibnamefont
  {Rubilar}},\ }\bibfield  {title} {\bibinfo {title} {Weak gravitational field
  effects on large-scale optical interferometric bell tests},\ }\href
  {https://doi.org/10.1088/1361-6382/ab8a60} {\bibfield  {journal} {\bibinfo
  {journal} {Classic. Quantum Grav.}\ }\textbf {\bibinfo {volume} {37}},\
  \bibinfo {pages} {195001} (\bibinfo {year} {2020})}\BibitemShut {NoStop}%
\bibitem [{\citenamefont {Brady}\ and\ \citenamefont {Haldar}(2021)}]{RelHOM}%
  \BibitemOpen
  \bibfield  {author} {\bibinfo {author} {\bibfnamefont {A.~J.}\ \bibnamefont
  {Brady}}\ and\ \bibinfo {author} {\bibfnamefont {S.}~\bibnamefont {Haldar}},\
  }\bibfield  {title} {\bibinfo {title} {Frame dragging and the hong-ou-mandel
  dip: Gravitational effects in multiphoton interference},\ }\href
  {https://doi.org/10.1103/PhysRevResearch.3.023024} {\bibfield  {journal}
  {\bibinfo  {journal} {Phys. Rev. Res.}\ }\textbf {\bibinfo {volume} {3}},\
  \bibinfo {pages} {023024} (\bibinfo {year} {2021})}\BibitemShut {NoStop}%
\bibitem [{\citenamefont {Roura}(2020)}]{Roura}%
  \BibitemOpen
  \bibfield  {author} {\bibinfo {author} {\bibfnamefont {A.}~\bibnamefont
  {Roura}},\ }\bibfield  {title} {\bibinfo {title} {Gravitational redshift in
  quantum-clock interferometry},\ }\href
  {https://doi.org/10.1103/PhysRevX.10.021014} {\bibfield  {journal} {\bibinfo
  {journal} {Phys. Rev. X}\ }\textbf {\bibinfo {volume} {10}},\ \bibinfo
  {pages} {021014} (\bibinfo {year} {2020})}\BibitemShut {NoStop}%
\bibitem [{\citenamefont {Dickerson}\ \emph {et~al.}(2013)\citenamefont
  {Dickerson}, \citenamefont {Hogan}, \citenamefont {Sugarbaker}, \citenamefont
  {Johnson},\ and\ \citenamefont {Kasevich}}]{Dickerson2013}%
  \BibitemOpen
  \bibfield  {author} {\bibinfo {author} {\bibfnamefont {S.~M.}\ \bibnamefont
  {Dickerson}}, \bibinfo {author} {\bibfnamefont {J.~M.}\ \bibnamefont
  {Hogan}}, \bibinfo {author} {\bibfnamefont {A.}~\bibnamefont {Sugarbaker}},
  \bibinfo {author} {\bibfnamefont {D.~M.~S.}\ \bibnamefont {Johnson}},\ and\
  \bibinfo {author} {\bibfnamefont {M.~A.}\ \bibnamefont {Kasevich}},\
  }\bibfield  {title} {\bibinfo {title} {Multiaxis inertial sensing with
  long-time point source atom interferometry},\ }\href
  {https://doi.org/10.1103/PhysRevLett.111.083001} {\bibfield  {journal}
  {\bibinfo  {journal} {Phys. Rev. Lett.}\ }\textbf {\bibinfo {volume} {111}},\
  \bibinfo {pages} {083001} (\bibinfo {year} {2013})}\BibitemShut {NoStop}%
\bibitem [{\citenamefont {Kovachy}\ \emph {et~al.}(2015)\citenamefont
  {Kovachy}, \citenamefont {Asenbaum}, \citenamefont {Overstreet},
  \citenamefont {Donnelly}, \citenamefont {Dickerson}, \citenamefont
  {Sugarbaker}, \citenamefont {Hogan},\ and\ \citenamefont
  {Kasevich}}]{Kovachy2015}%
  \BibitemOpen
  \bibfield  {author} {\bibinfo {author} {\bibfnamefont {T.}~\bibnamefont
  {Kovachy}}, \bibinfo {author} {\bibfnamefont {P.}~\bibnamefont {Asenbaum}},
  \bibinfo {author} {\bibfnamefont {C.}~\bibnamefont {Overstreet}}, \bibinfo
  {author} {\bibfnamefont {C.~A.}\ \bibnamefont {Donnelly}}, \bibinfo {author}
  {\bibfnamefont {S.~M.}\ \bibnamefont {Dickerson}}, \bibinfo {author}
  {\bibfnamefont {A.}~\bibnamefont {Sugarbaker}}, \bibinfo {author}
  {\bibfnamefont {J.~M.}\ \bibnamefont {Hogan}},\ and\ \bibinfo {author}
  {\bibfnamefont {M.~A.}\ \bibnamefont {Kasevich}},\ }\bibfield  {title}
  {\bibinfo {title} {Quantum superposition at the half-meter scale},\ }\href
  {https://doi.org/10.1038/nature16155} {\bibfield  {journal} {\bibinfo
  {journal} {Nature}\ }\textbf {\bibinfo {volume} {528}},\ \bibinfo {pages}
  {530} (\bibinfo {year} {2015})}\BibitemShut {NoStop}%
\bibitem [{\citenamefont {{VLBAI at HITec}}()}]{Hannover}%
  \BibitemOpen
  \bibfield  {author} {\bibinfo {author} {\bibnamefont {{VLBAI at HITec}}},\
  }\href@noop {} {\ }\Eprint
  {https://arxiv.org/abs/https://www.hitec.uni-hannover.de/en/large-scale-equipment/atomic-fountain}
  {https://www.hitec.uni-hannover.de/en/large-scale-equipment/atomic-fountain}
  \BibitemShut {NoStop}%
\bibitem [{\citenamefont {Bernardini}\ \emph {et~al.}(2004)\citenamefont
  {Bernardini}, \citenamefont {Leo},\ and\ \citenamefont
  {Rotelli}}]{Bernardini}%
  \BibitemOpen
  \bibfield  {author} {\bibinfo {author} {\bibfnamefont {A.}~\bibnamefont
  {Bernardini}}, \bibinfo {author} {\bibfnamefont {S.}~\bibnamefont {Leo}},\
  and\ \bibinfo {author} {\bibfnamefont {P.}~\bibnamefont {Rotelli}},\
  }\bibfield  {title} {\bibinfo {title} {Above potential barrier diffusion},\
  }\href {https://doi.org/10.1142/S0217732304015877} {\bibfield  {journal}
  {\bibinfo  {journal} {Mod. Phys. Lett. A}\ }\textbf {\bibinfo {volume}
  {19}},\ \bibinfo {pages} {2717} (\bibinfo {year} {2004})}\BibitemShut
  {NoStop}%
\bibitem [{\citenamefont {Norsen}\ \emph {et~al.}()\citenamefont {Norsen},
  \citenamefont {Lande},\ and\ \citenamefont {McKagan}}]{Norsen}%
  \BibitemOpen
  \bibfield  {author} {\bibinfo {author} {\bibfnamefont {T.}~\bibnamefont
  {Norsen}}, \bibinfo {author} {\bibfnamefont {J.}~\bibnamefont {Lande}},\ and\
  \bibinfo {author} {\bibfnamefont {S.~B.}\ \bibnamefont {McKagan}},\
  }\href@noop {} {\bibinfo {title} {How and why to think about scattering in
  terms of wave packets instead of plane waves}},\ \Eprint
  {https://arxiv.org/abs/0808.3566} {arXiv:0808.3566} \BibitemShut {NoStop}%
\end{thebibliography}%

\end{document}